\documentclass{article}
\usepackage{graphicx}  
\usepackage{amsmath}   
\usepackage[compress]{cite}
\usepackage{amssymb}   
\usepackage{bm} 
\usepackage{dcolumn}
\usepackage{color}
\usepackage{float}
\usepackage{mathrsfs}
\usepackage{amsfonts}
\usepackage{varioref}
\usepackage[utf8]{inputenc}
\RequirePackage[colorlinks,citecolor=blue,urlcolor=magenta,linkcolor=blue]{hyperref}
\def\gr{general relativity}

\def\GB{Gauss-Bonnet }
\labelformat{section}{Section #1} 
\labelformat{subsection}{Section #1} 
\labelformat{subsubsection}{Section #1}
\labelformat{subsubsubsection}{Section #1}
\labelformat{equation}{Eq.~(#1)} 
\labelformat{figure}{Fig.~#1} 
\labelformat{subfigure}{Fig.~\thefigure#1} 
\labelformat{table}{Tab.~#1} 
\labelformat{appendix}{Appendix #1}
\title{Inflation driven by Einstein-Gauss-Bonnet gravity}

\author{Sumanta Chakraborty~\footnote{sumantac.physics@gmail.com}, Tanmoy Paul~\footnote{pul.tnmy9@gmail.com} and Soumitra SenGupta~\footnote{tpssg@iacs.res.in}\\
{\small{Department of Theoretical Physics,
Indian Association for the Cultivation of Science, Kolkata-700032, India}}}
\begin{document}
  
\maketitle
\begin{abstract}
We have explicitly demonstrated that scalar coupled Gauss-Bonnet gravity in four dimension can have non-trivial effects on the early inflationary stage of our universe. In particular, we have shown that the scalar coupled Gauss-Bonnet term alone is capable of driving the inflationary stages of the universe \emph{without} incorporating slow roll approximation, while remaining compatible with the current observations. Subsequently, to avoid the instability of the tensor perturbation modes we have introduced a self-interacting potential for the inflaton field and have shown that in this context as well it is possible to have inflationary scenario. Moreover it turns out that presence of the Gauss-Bonnet term is incompatible with the slow roll approximation and hence one must work with the field equations in the most general context. Finally, we have shown that the scalar coupled Gauss-Bonnet term attains smaller and smaller values as the universe exits from inflation. Thus at the end of the inflation the universe makes a smooth transition to Einstein gravity.
\end{abstract}
\section{Introduction}

General relativity describes the gravitational interaction in its simplest form. Since viability of any theory is based on its falsifiable predictions and consistency with existing observations, one can safely argue that \gr\ is the most viable theory of gravitation till date. This is mainly due to the fact that so far \gr\ has passed the experimental tests with flying colours \cite{will1993theory,carroll2004spacetime,gravitation,MTW}. However, as it is necessary for advancement of theoretical sciences, despite its enormous successes, \gr\ is also riddled with many open questions. These are scattered across various length scales and include the inflationary epoch and big bang singularity in the context of early universe cosmology \cite{Guth:1980zm,Starobinsky:1980te,Linde:1981mu,Linde:1982zj,Riess:1998cb,
Perlmutter:1998np,Rubakov:1983bz,Peebles:2002gy,Carroll:2000fy,Padmanabhan:2002ji}, which we will concentrate on in this work. In this particular context there exists several issues among which, flatness of the universe at a large scale, uniformity of the temperature of Cosmic Microwave Background in super-horizon scales are some of the important ones. These problems are believed to be answered in one way or another by the introduction of various inflationary models of our universe \cite{Guth:1980zm,Starobinsky:1980te,Linde:1981mu,Linde:1982zj,Abbott:1982hn,Linde:1983gd}. According to the standard inflationary paradigm, in the very early stages the universe went through an exponentially accelerating expansion, which later on starts to decelerate and makes path for the standard cosmological epochs. One of the most popular attempt to achieve the same is by considering a scalar field with a self-interacting potential sourcing gravity and assuming that the scalar field satisfies the ``slow-roll'' condition (i.e., kinetic energy of the scalar field is very much less than the potential energy) \cite{Dodelson:2003ft,Turok:2002yq,Lyth:1993eu,Liddle:1999mq,Brandenberger:1999sw,Guth:2000ka,
Lidsey:1995np,Burgess:2001vr} (however also see \cite{Padmanabhan:1988ji,Padmanabhan:1988se}). Therefore most of the inflationary paradigms are driven by a scalar field with a non-trivial self-interacting potential in Einstein gravity.

A natural pathway through which such a scalar field can enter the gravitational dynamics at the early universe is through the coupling of the field with the Gauss-Bonnet term. The \GB term is the first non-trivial higher curvature correction to the Einstein-Hilbert action \cite{Zwiebach:1985uq,Gross:1986mw,Dadhich:2008df,Padmanabhan:2013xyr}, leading to second order field equations and hence avoiding the Ostrogradsky instability \cite{Woodard:2015zca}. Even though the \GB term alone, in the context of four dimensional physics, does not contribute to the gravitational field equations, the scalar coupling makes the \GB term (and hence the field equations) non-trivial. Some aspects of this scalar coupled \GB gravity in the context of early universe physics has been explored in \cite{Satoh:2008ck,Guo:2010jr,Jiang:2013gza,Koh:2014bka,Kanti:2015pda,Kanti:2015dra,Lahiri:2016jqv,vandeBruck:2016xvt,Koh:2016abf,Sberna:2017xqv,Fomin:2017vae,vandeBruck:2017voa} (for a set of earlier works in other alternative theories in the similar spirit, see \cite{Nojiri:2010wj,Sotiriou:2008rp,DeFelice:2010aj,Nojiri:2007as,
Nojiri:2003ft,Nojiri:2017ncd,Oikonomou:2018npe,Chakraborty:2014xla,
Chakraborty:2015bja,Capozziello:1996xg,Sotiriou:2006hs,Catena:2006bd,Chakraborty:2016ydo,
Chakraborty:2016gpg,Paul:2018kdq,Karam:2018squ,Sami:2017nhw,Barrow:1988,Ellis:1991,Cognola:2005de,Bamba:2008ja,Appleby:2009uf,
Sebastiani:2015kfa,Banerjee:2017lxi,Das:2017jrl,
Chakraborty:2015wma,Chakraborty:2015taq,Chakraborty:2017zep,Cognola:2006eg,Nojiri:2005jg,Antoniadis:1993jc,Kanti:1995vq,Kanti:1997br,Charmousis:2002rc,
Binetruy:2002ck,Germani:2002pt,Gravanis:2002wy,
Nojiri:2005am,Leith:2007bu,Deser:2007jk,Bamba:2014mya,
vandeBruck:2015xpa,Sotiriou:2013qea,Hees:2017aal,Antoniou:2017acq,Charmousis:2014mia,
Chakraborty:2016lxo,Banerjee:2017hzw,Mukherjee:2017fqz,Pirtskhalava:2015nla,
Banerjee:2016hom,Banerjee:2017jyb,Bhattacharya:2016naa,Banerjee:2018yyi}). Below we provide a brief discussion on the results obtained in these works.

The inflationary paradigm has been explored in \cite{Kanti:2015pda,Kanti:2015dra} \emph{only} in the context of scalar coupled \GB gravity, excluding the Einstein term. While in \cite{vandeBruck:2017voa,vandeBruck:2016xvt}, even though the Einstein term was essential, the self-interacting potential \emph{itself} governs the inflation, having no effect of the \GB term. On the other hand, in \cite{Satoh:2008ck,Guo:2010jr,Jiang:2013gza,Koh:2014bka} both the self-interacting potential as well as the Gauss-Bonnet coupling for the inflaton field has been considered, \emph{but} in the context of slow-roll approximation (see also \cite{Koh:2014bka,Koh:2016abf,Sberna:2017xqv,Fomin:2017vae,Aghanim:2015xee,Ade:2015xua}). Thus non-trivial effects of the scalar coupled \GB term in the Einstein-Hilbert action, in \emph{absence} of self-interacting scalar potential in the context of inflationary paradigm has not been explored before. Besides, even when the self-interacting potential is added to the action, the relevant consequences of \emph{not} incorporating the slow-roll approximation in the inflationary paradigm deserves attention. 
 
In this paper, we would like to fill this gap by describing the inflationary paradigm with the help of scalar coupled \GB term in the Einstein-Hilbert action, without any self-interacting potential for the scalar field. We will demonstrate that such a scalar coupled \GB term alone (of course, in presence of the Ricci scalar) is capable of driving the exponential expansion of the early universe and also leads to an exit from the same, while remaining consistent with the current observations. However instability of the tensor perturbation in scalar coupled \GB gravity forced us to introduce the self-interacting potential for the scalar field. In this context as well, \emph{without} assuming the slow-roll approximation for the scalar field, we can trace over the whole inflationary epoch, which shows an initial de Sitter phase and a final deceleration phase effecting exit from the inflation.  

This paper is organized as follows --- In \ref{Sec_EGB_Int} we present the field equations associated with the scalar coupled Einstein-Gauss-Bonnet gravity in the context of cosmology. Subsequently, in \ref{Inf_self_int} we demonstrate that it is indeed possible to have inflationary scenario without the self-interacting potential term, while remaining consistent with observations. A possible source of instability of this model has also been presented in \ref{instab_GB}. Finally we have introduced a scalar potential and have demonstrated that the theory supports two different sets of analytic solutions for different choices of the scalar field potential and coupling function of scalar field with the \GB term in \ref{Inf_GB_Pot}. We finish the paper by providing some concluding remarks and future directions of exploration. 

\emph{Notations and Conventions} --- Throughout this paper Greek indices have been used to represent four-dimensional quantities. The fundamental constants $c$ and $\hbar$ have been set to unity, while the Newton's constant $G$ has been kept throughout. We have adopted the mostly positive signature.  
\section{Scalar coupled Einstein-Gauss-Bonnet gravity}\label{Sec_EGB_Int}

We consider a scalar coupled theory of gravity involving higher curvature terms, in which the scalar field is non-minimally coupled to the Gauss-Bonnet invariant $\mathcal{G}=R^{2}-4R_{\mu \nu}R^{\mu \nu}+R^{\alpha \beta \rho \sigma}R_{\alpha \beta \rho \sigma}$ in four dimensional spacetime. Therefore in the most general setting, the action for the scalar coupled Einstein-Gauss-Bonnet gravity consists of four terms --- (a) The Ricci scalar, (b) The Gauss-Bonnet invariant coupled to an arbitrary function of the scalar field, (c) kinetic term of the scalar field and finally (d) a self-interaction term for the scalar field, such that
\begin{align}\label{action}
\mathcal{A}=\int d^4x \sqrt{-g}\bigg[\frac{1}{16\pi G}\Big\{R-\xi(\Phi)\mathcal{G}\Big\}
-\frac{1}{2}g^{\mu\nu}\partial_{\mu}\Phi\partial_{\nu}\Phi -V(\Phi)\bigg]~,
\end{align}
where $R$ is the Ricci scalar obtained from the metric $g_{\mu\nu}$, $\Phi$ is the scalar field under consideration and $\mathcal{G}$, defined earlier, is the Gauss-Bonnet invariant. The non-topological character of the Gauss-Bonnet term in the above action is ensured by the coupling function between the scalar field and the \GB term, symbolized by $\xi(\Phi)$. One possible origin of the term $\xi(\Phi)\mathcal{G}$ is from the compactification of a higher dimensional spacetime to an effective four dimensional description, where $\Phi$ plays the role of the radion field \cite{Amendola:2005cr}. 

Variation of the above action, presented in \ref{action}, with respect to the metric and the scalar field results into the following field equations for gravity and the scalar field individually,
\begin{align}
G_{\mu \nu}&+\frac{1}{2}g_{\mu \nu}~\xi(\Phi)\mathcal{G}
-2\xi(\Phi)\Big[RR_{\mu\nu}-2R_{\mu \rho}R^{\rho}_{\nu}
+R_{\mu}^{~\rho\sigma\tau}R_{\nu\rho\sigma\tau}-2R_{\mu\rho\nu\sigma}R^{\rho\sigma}\Big]
\nonumber
\\ 
&+2\{\nabla_{\mu}\nabla_{\nu}\xi(\Phi)\}R-2g_{\mu\nu}\{\nabla^2\xi(\Phi)\}R 
-4\{\nabla_{\rho}\nabla_{\mu}\xi(\Phi)\}R_{\nu}^{\rho} 
-4\{\nabla_{\rho}\nabla_{\nu}\xi(\Phi)\}R^{\rho}_{\mu}
\nonumber
\\ 
&+4\{\nabla^2\xi(\Phi)\}R_{\mu\nu} 
+4g_{\mu\nu}\{R^{\rho\sigma}\nabla_{\rho}\nabla_{\sigma}\xi(\Phi)\} 
+4\{\nabla^{\rho}\nabla^{\sigma}\xi(\Phi)\}R_{\mu\rho\nu\sigma} 
\nonumber
\\
&\hspace{3cm}=8\pi G\Big[\nabla_{\mu}\Phi\nabla_{\nu}\Phi 
-g_{\mu\nu}\left\{\frac{1}{2}\nabla_{\rho}\Phi\nabla^{\rho}\Phi+V(\Phi)\right\}\Big];
\label{gravitational_equation}
\\
\square\Phi&-\left(\frac{\partial \xi}{\partial \Phi}\right)\frac{\mathcal{G}}{16\pi G}-\frac{\partial V}{\partial \Phi}=0~.
\label{scalar_equation}
\end{align}
As expected, the gravitational field equations do not contain more than second order derivatives of the metric and hence is intrinsically ghost free. We will apply the above general analysis in the context of inflationary paradigm, where the higher curvature effects are supposed to be important \cite{Kanti:2015pda,vandeBruck:2017voa}. 

In the context of inflationary paradigm it is customary to choose the background spacetime to be described by a homogeneous, isotropic and spatially flat metric, which takes the following form,
\begin{equation}\label{metric_ansatz}
ds^{2}=-dt^{2}+a^{2}(t)\big\{dx^{2}+dy^{2}+dz^{2}\big\}~,
\end{equation}
where the scale factor $a(t)$ solely governs evolution of the spacetime structure. For such a metric, the expression for the Ricci scalar $R$ and the \GB invariant $\mathcal{G}$ can be easily computed, which results into,
\begin{align}\label{Inv_R_G}
R=6\left(2H^{2}+\dot{H}\right);\qquad \mathcal{G}=24H^{2}\left(H^{2}+\dot{H}\right)~,
\end{align}
with $H=\dot{a}/a$ and `dot' denotes derivative of the respective quantity with respect to time. In order to be consistent with the symmetry of the background spacetime it is necessary that the inflaton field be dependent on the time coordinate alone, i.e., $\Phi=\Phi(t)$. Finally, using the expressions for the Ricci scalar and the \GB invariant from \ref{Inv_R_G}, along with the Ricci and Riemann tensor for the spacetime metric presented in \ref{metric_ansatz}, the field equations \emph{in absence of potential} can be simplified, leading to,
\begin{align}
3H^{2}-12H^{3}\dot{\xi}&=8\pi G\left(\frac{1}{2}\dot{\Phi}^{2}\right);
\label{gr_equation_temporal_part}
\\
2\dot{H}-4H^2\Big[\ddot{\xi}-H\dot{\xi}&+2\frac{\dot{H}}{H}\dot{\xi}\Big]=-8\pi G~\dot{\Phi}^2;
\label{gr_equation_spatial_part}
\\
\ddot{\Phi}+3H\dot{\Phi}+\frac{12H^{2}}{8\pi G}\Big(H^{2}&+\dot{H}\Big)\frac{\partial \xi}{\partial \Phi}=0~.
\label{scalar_field_equation}
\end{align}
It is evident that due to the presence of the \GB term, cubic as well as quartic powers of $H(t)$ appear in the above field equations. Further due to Bianchi identity and conservation of matter energy momentum tensor, all the three field equations presented above are not independent, but one of them can be derived from the other two. For example, one can derive \ref{gr_equation_spatial_part} by differentiating \ref{gr_equation_temporal_part} with respect to the time coordinate and then using \ref{scalar_field_equation} to replace $\ddot{\Phi}$. Similarly, using \ref{gr_equation_temporal_part} and \ref{gr_equation_spatial_part} it is possible to derive \ref{scalar_field_equation} as well. 

The best way to describe the inflationary paradigm is through the slow-roll approximation imposed on the scalar field, which requires $\dot{\Phi}^{2}\ll \dot{\Phi}$ and $\ddot{\Phi}\ll \dot{\Phi}$. Under these approximations the gravitational field equation for the scale factor $a(t)$, presented in \ref{gr_equation_temporal_part}, simplifies considerably and it becomes possible to solve for $\dot{\Phi}$, yielding
\begin{align}\label{slow_phidot}
\dot{\Phi}=\frac{1}{4H}\left(\frac{\partial \xi}{\partial \Phi}\right)^{-1}~.
\end{align}
On the other hand, the field equation for the scalar field, as in \ref{scalar_field_equation}, under the slow-roll approximation result into $3H\dot{\Phi}$ to be proportional to $H^{2}(\dot{H}+H^{2})(\partial \xi/\partial \Phi)$. Therefore, by substituting the expression for $\dot{\Phi}$ from \ref{slow_phidot} one immediately obtains the following result for $\dot{H}+H^{2}$,
\begin{align}
\dot{H}+H^{2}=-\frac{\pi G}{2H^{2}}\left(\frac{\partial \xi}{\partial \Phi}\right)^{-2}~.
\end{align}
The above expression explicitly shows that $\ddot{a}/a=\dot{H}+H^{2}$ is a negative quantity, since neither $H$, nor $(\partial \xi/\partial \Phi)$ are imaginary. The above result ensures that \emph{under slow-roll approximation}, it is \emph{impossible} to arrive at an inflationary solution for our universe in this context. One would therefore tend to introduce a self-interacting potential term to achieve the desired slow-roll inflation. However, we will show that even in the absence of such a self-interacting potential one can still have inflationary solutions compatible with current observations \emph{without} going into the slow-roll approximation. This is what we will elaborate in the next section.
\section{Inflation without a self-interacting potential}\label{Inf_self_int}

This section is devoted to the study of inflationary paradigm in the absence of self-interacting potential, but with a non-minimal coupling of the scalar field with \GB invariant. As we have argued before, the slow-roll approximation can not lead to an inflationary paradigm and hence we would now like to go beyond this approximation. To set the stage, let us first ask whether it is possible to have any solution for $\xi(\Phi)$ with constant Hubble parameter in absence of potential term for the inflaton field. If this can be achieved then only one can proceed further and try to obtain a complete inflationary scenario which is compatible with the current observational constraints. 
\subsection{Possibility for constant Hubble parameter}\label{Eternal_Inflation}

In this section we will concentrate on the possibility of having constant Hubble parameter  (i.e., $H(t) = H_{0}=\textrm{constant}$), which is consistent even without the potential term for the inflaton field. In other words, we have to use the fact that Hubble parameter is constant, in the field equations for gravity as well as the scalar field and then inspect whether a non-trivial solution for $\xi(\Phi)$ can be obtained. Keeping this in mind, we rewrite  \ref{gr_equation_temporal_part} and \ref{gr_equation_spatial_part} in the following manner,
\begin{align}
3H_{0}^{2}-12H_{0}^{3}\dot{\xi}=8\pi G\left(\frac{1}{2}\dot{\Phi}^{2}\right);
\label{gr_equation_temporal_part1}
\\
4H_{0}^{2}\left(\ddot{\xi}-H_{0}\dot{\xi}\right)=8\pi G~ \dot{\Phi}^{2}~.
\label{gr_equation_spatial_part2}
\end{align}
Given the above equations one can eliminate the $\dot{\Phi}^2$ term from both of them and obtain the following second order differential equation for $\xi(t)$, $2\ddot{\xi}+10H_{0}\dot{\xi}-3=0$. It is straightforward to solve for $\xi(t)$ given the above equation and it turns out to be,
\begin{align}
\xi(t)=\frac{1}{5H_{0}}\Bigg[\frac{3}{2}t+Ae^{-5H_0t}\bigg]+B~,
\label{solution_of_xi_1}
\end{align}
where $A$ and $B$ are constants of integration. The above solution for $\xi(t)$ when substituted in \ref{gr_equation_temporal_part1} immediately leads to the following first order differential equation for $\Phi(t)$,
\begin{align}
8\pi G~\dot{\Phi}^2 =24AH_0^{3}e^{-5H_{0}t}-\frac{6H_{0}^{2}}{5}~.
\label{differential_equation_for_phi_1}
\end{align}
The above equation can be readily integrated yielding the following solution for the inflaton field $\Phi(t)$ as,
\begin{align}
\sqrt{8\pi G}~\Phi(t) = \frac{2\sqrt{6}}{5\sqrt{5}} \Bigg[\tan^{-1}\bigg(\sqrt{20AH_{0}e^{-5H_0t}-1}\bigg)-\sqrt{20AH_{0}e^{-5H_{0}t}- 1}\bigg]~.
\label{solution_of_phi_1}
\end{align}
Note that in order to have a real solution it is of utmost importance to have $A>0$, otherwise the term within the square root will turn negative. For $A>0$ one will have non-trivial time evolution for the inflaton field as well as for the coupling $\xi(\Phi)$ as evident from \ref{solution_of_phi_1}. Therefore, the scalar coupled Einstein-Gauss-Bonnet gravity without any self-interaction term for the scalar field is capable of producing exponential expansion of the universe. However there is one major shortcoming of the above result, namely it does not predict when the inflation will end. It is easy to determine from \ref{solution_of_phi_1} that after a time $t\equiv t_{\rm f}=(1/5H_{0})\ln (20AH_{0})$ the $H=H_{0}=\textrm{constant}$ solution is no longer valid. However the model can not explain any natural mechanism to exit from the inflation before $t=t_{f}$. Therefore, in order to describe the inflationary era of the early universe consistently it is necessary for the inflation to end and the duration of 
inflation, represented by the number of e-foldings, must be in consonance with the recent Planck observations. 
\subsection{Inflation with an exit}

In this section, we will demonstrate that it is indeed possible to have a proper inflationary phase in the early universe described by the scalar coupled Einstein-Gauss-Bonnet gravity \emph{without} any self-interacting scalar potential. For this purpose, we first consider the simpler scenario presented in \ref{Eternal_Inflation}. As evident from \ref{solution_of_xi_1} and \ref{solution_of_phi_1} it is not possible to write $\xi=\xi(\Phi)$ in a closed form, since the solution for $\Phi(t)$ is a transcendental equation. Therefore, in the more general context we should not expect a simple closed form expression for the coupling function $\xi(\Phi)$. 

Given this difficulty, we will employ the well known reconstruction scheme in order to arrive at a viable inflationary model in the present context \cite{Carloni:2010ph,Nojiri:2009xh,Nojiri:2009kx,Chakraborty:2016ydo}. As a first step of this reconstruction method, we start with a particular ansatz for the time dependence of the Hubble parameter $H(t)$ and ensure that it is indeed consistent with the observational constraints, i.e., it predicts correct value of the tensor to scalar ratio and the scalar spectral index. Given the Hubble parameter, one can immediately eliminate $\dot{\Phi}$ between \ref{gr_equation_temporal_part} and \ref{gr_equation_spatial_part} respectively. This results into the following second order differential equation for $\xi(t)$
\begin{align}\label{Diff_Eq_xi}
\ddot{\xi}+\left(5H+2\frac{\dot{H}}{H}\right)\dot{\xi}-\left(\frac{\dot{H}}{2H^{2}}+\frac{3}{2}\right)=0
\end{align}
One can integrate the above equation by multiplying both sides by the integrating factor, which reads,
\begin{align}
\textrm{Integrating Factor}=\exp \left[\int dt \left(2\frac{\dot{H}}{H}+5H\right)\right]
\equiv e^{P}
\end{align}
Therefore multiplying both sides of \ref{Diff_Eq_xi} by the integrating factor $e^{P}$
one can immediately integrate the above second order differential equation for $\xi(t)$  yielding,
\begin{align}
\dot{\xi}(t)=e^{-P(t)}\int dt'e^{P(t')}\left\{\frac{\dot{H}(t')}{2H(t')^{2}}+\frac{3}{2}\right\}
+C_{1}e^{-P(t)}
\end{align}
Finally integrating the above differential equation once again we arrived at, 
\begin{align}\label{xi_sol}
\xi(t)=\int dt e^{-P}\int dt'e^{P(t')}\left\{\frac{\dot{H}(t')}{2H(t')^{2}}+\frac{3}{2}\right\}+C_{1}\int dt e^{-P(t)}+C_{2}
\end{align}
where $C_{1}$ and $C_{2}$ are constants of integration. Thus having derived the coupling function $\xi(t)$ the time evolution of the scalar field follows from the following differential equation 
\begin{align}\label{phi_dot}
4\pi G~\dot{\Phi}^{2}=3H^{2}-12H^{3}\Bigg[e^{-P(t)}\int dt'e^{P(t')}\left\{\frac{\dot{H}(t')}{2H(t')^{2}}+\frac{3}{2}\right\}
+C_{1}e^{-P(t)}\Bigg]
\end{align}
At this stage, it deserves mentioning that at initial stages of the inflation, the Hubble parameter is almost constant and hence one may assume $H=H_{0}=\textrm{constant}$. This situation has already been discussed in \ref{Eternal_Inflation} and one may derive the relevant results by setting $\dot{H}=0$ in \ref{xi_sol} and \ref{phi_dot} respectively. 

So far, we have kept our discussion completely general and have not specified any particular choice for the Hubble parameter $H(t)$. The choice for the Hubble parameter cannot be arbitrary, as it must satisfy the following condition: at the onset of inflation the Hubble parameter must take nearly constant values. Further keeping in mind that a natural exit from the inflationary dynamics is necessary, here we propose a time dependent ansatz for the Hubble parameter as follows:
\begin{align}\label{solution_of_H_2}
H(t)=\bigg[c - d\left(t-t_{*}\right)\bigg]^{\alpha}~,
\end{align}
where $c$, $d$ and $\alpha$ are the free parameters of the theory. The time scale $t_{*}$ is assumed to represent the onset of inflation and as evident from the above ansatz, for $t\sim t_{*}$ the Hubble parameter is almost constant with $H\sim c^{\alpha}$. Therefore at the beginning of inflation we have a very small value for $\dot{H}$ which will subsequently grow and will be order of the Hubble parameter requiring the inflation to end. Therefore, we may introduce a dimensionless variable $\epsilon (t)$ as $-\dot{H}/H^{2}$. From the previous discussion it is clear that $\epsilon \ll 1$ at the onset of inflation, while $\epsilon \sim 1$ as the inflation ends. This ensures that $\dot{H}+H^{2}>0$ throughout the course of inflation. Using the explicit form of the Hubble parameter $H(t)$ from \ref{solution_of_H_2}, the parameter $\epsilon(t)$ can be computed such that,
\begin{align}
\epsilon(t)=\alpha d\Big\{c - d(t-t_{*})\Big\}^{-\alpha -1}~.
\label{epsilon}
\end{align}
Since the Hubble parameter and hence $c^{\alpha}$ is much larger than unity it follows that for $t\sim t_{*}$, $\epsilon$ is much smaller compared to unity. The above expression of $\epsilon(t)$ can also be used to determine the end of inflation as well. For this we assume that the exit time of inflation, i.e., $t_{f}$ is being determined by the equation $\epsilon(t_{\rm f})=1$. This immediately leads to the following expression for $\Delta t=t_{f}-t_{*}$, corresponding to the duration of inflation as,
\begin{align}
\Delta t \equiv t_{\rm f}-t_{*}=\frac{1}{d}\left\{c-\left(\alpha d\right)^{1/(1+\alpha)}\right\}~.
\label{duration}
\end{align}
Moreover, \ref{epsilon} clearly reveals that $\epsilon(t)$ remains less than unity for $t_{*}<t<t_{\rm f}$. Therefore the above ansatz for Hubble parameter can describe the evolution of the universe during inflationary epoch quite well. The parameter $\epsilon$ starts from a small value at $t\sim t_{*}$ and then grows to become order unity as $t\sim t_{\rm f}$ and then the universe exits from inflation. The above analysis also enables us to estimate the values of the parameters, namely $c$ and $d$. This can be obtained by requiring the above expression for the Hubble parameter in \ref{solution_of_H_2} to remain valid till the end of inflation. This requires $c/d>\Delta t$, which by using the duration of inflation, demands  $c/d>10^{-7}\textrm{GeV}^{-1}$. This also suggests that $t_{*}$ should have a value $\sim 10^{-11}\textrm{GeV}^{-1}$. In the present context we have chosen the ratio $c/d\sim 10^{-3}\textrm{GeV}^{-1}$ so that the Hubble parameter remain valid throughout the duration of the inflation. Note that the time scale $t_{\rm e}$ for which $(t_{\rm e}-t_{*})>c/d$ will never arise, since this would correspond to a scenario much after the end of inflation, where the above solution is no longer valid. 

In order to be compatible with precision observations associated with the inflationary paradigm \cite{Aghanim:2015xee,Ade:2015xua}, it is crucial to compute various parameters of experimental interest, for which the number of e-foldings in the present context reads 
\begin{align}
N\equiv \int_{t_{*}}^{t_{\rm f}} H(t) dt
=\frac{c^{\alpha + 1}}{d(\alpha + 1)}-\frac{\left\{c - d\left(t_{\rm f}-t_{*}\right)\right\}^{\alpha + 1}}{d(\alpha + 1)}~.
\label{e_folding_1}
\end{align}
In order to arrive at the last line, the solution for $H(t)$ from \ref{solution_of_H_2} has been used in order to perform the integral in the definition of the number of e-foldings. Substitution of the time span for inflation from \ref{duration} further simplifies the above expression and one finally obtains the number of e-foldings as follows:
\begin{align}
N=\frac{c^{\alpha + 1}}{d(\alpha + 1)}-\frac{\alpha}{\alpha + 1}~.
\label{e_folding_3}
\end{align}
Having determined the number of e-foldings let us concentrate on the possible observables associated with this model. Before going into the details of computation, let us briefly recall what these observables essentially measures. The gravitational perturbation around the Friedman metric can be decomposed into three categories: scalar perturbations, vector perturbations and finally tensor perturbations. The vector perturbations generally die down and hence one normally considers the scalar and the tensor perturbations. Assuming the perturbations to be Gaussian one can encode all the information about the perturbation in the power spectral density, i.e., how much power is contained for each length scale or equivalently for each wave mode. From this it is immediate to compute the power spectrum, whose Logarithmic derivative with respect to the wave number provides the corresponding spectral index (also known as the tilt). The spectral index for scalar perturbation, known as $n_{s}$ and the ratio of power spectrum of the tensor perturbation and scalar perturbation, known as tensor-to-scalar ratio $r$ are the observables we will use. In absence of potential both the scalar spectral index and the tensor-to-scalar ratio can be written solely in terms of the parameter $\epsilon$ \cite{Sberna:2017nzp}, since all the corrections to them identically vanishes if the potential is set to zero. Given the above it turns out that the associated observables, namely the tensor to scalar ratio $r$ and the spectral index of curvature perturbation $n_{\rm s}$ can be determined using the number of e-foldings and parameter $\alpha$ appearing in the expression for Hubble parameter. Thus, using \ref{e_folding_3} and \ref{solution_of_H_2} the tensor to scalar ratio and the scalar spectral index becomes, 
\begin{align}
r&=16\epsilon(t_{*})=16~\Big[N\bigg(\frac{\alpha + 1}{\alpha}\bigg)+1\Big]^{-1};
\label{tensor_scalar_ratio}
\\
n_{\rm s}&=1-2\epsilon(t_{*})-\frac{\dot{\epsilon}}{H\epsilon}\bigg|_{t_{*}}
=1-\frac{(3\alpha + 1)}{(\alpha + 1)}\left\{N + \frac{\alpha}{\alpha + 1}\right\}^{-1}~.
\label{spectral_index}
\end{align}
In order to derive \ref{tensor_scalar_ratio} and \ref{spectral_index} respectively, we have used the expression for the number of e-foldings that has been obtained in \ref{e_folding_3}. From current precession cosmology one has the following bounds on the tensor to scalar ratio $r$ and the spectral index of curvature perturbation $n_{\rm s}$: $n_{\rm s}=0.968\pm 0.006$ and $r<0.14$ respectively. The above constraints essentially originate from the joint analysis of temperature cross correlations in the Cosmic Microwave Background and the weak gravitational lensing  obtained from Planck satellite \cite{Aghanim:2015xee,Ade:2015xua}. Using \ref{tensor_scalar_ratio} and 
\ref{spectral_index}, it can be easily shown that in order to have the theoretical estimates to be consistent with the observational results, $N$ and $\alpha$ should be equal to $60$ and $\frac{3}{5}$ respectively. Putting these values of $N$, $\alpha$ into \ref{tensor_scalar_ratio} and \ref{spectral_index}, we obtain the following numerical estimates for $r$ and $n_{\rm s}$ such that, $r=0.10$ and $n_{\rm s}=0.970$, which are well within the experimental bounds. Therefore the Hubble parameter as presented in \ref{solution_of_H_2} is indeed compatible with current observational bounds, provided the parameter $\alpha\simeq 3/5$. Thus using the reconstruction scheme we have been able to determine a suitable Hubble parameter, which we will use subsequently to determine the coupling function $\xi(\Phi)$. 

\begin{figure}[!h]
\centering
\includegraphics[width=3in,height=2in]{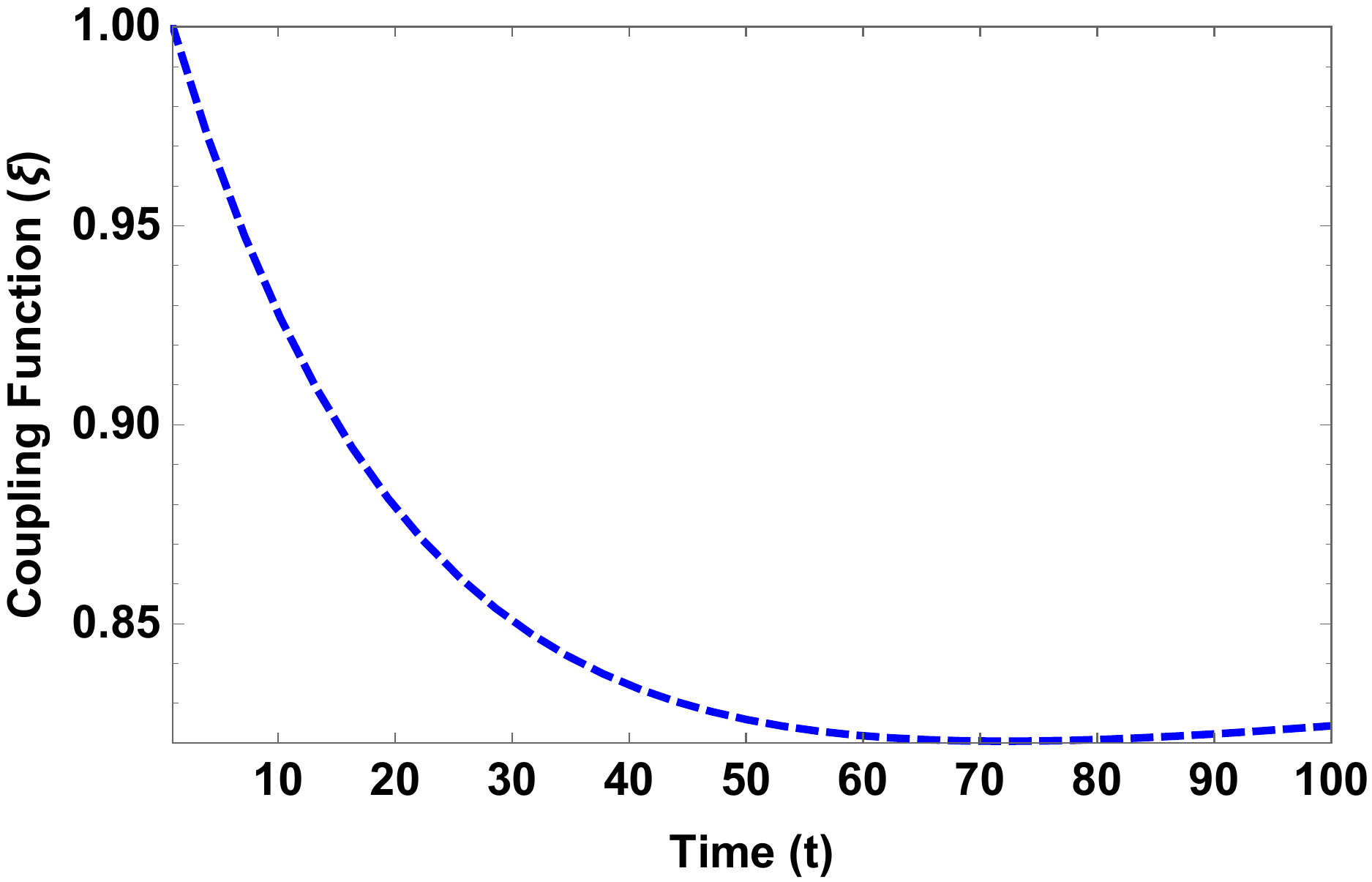}~~
\includegraphics[width=3in,height=2in]{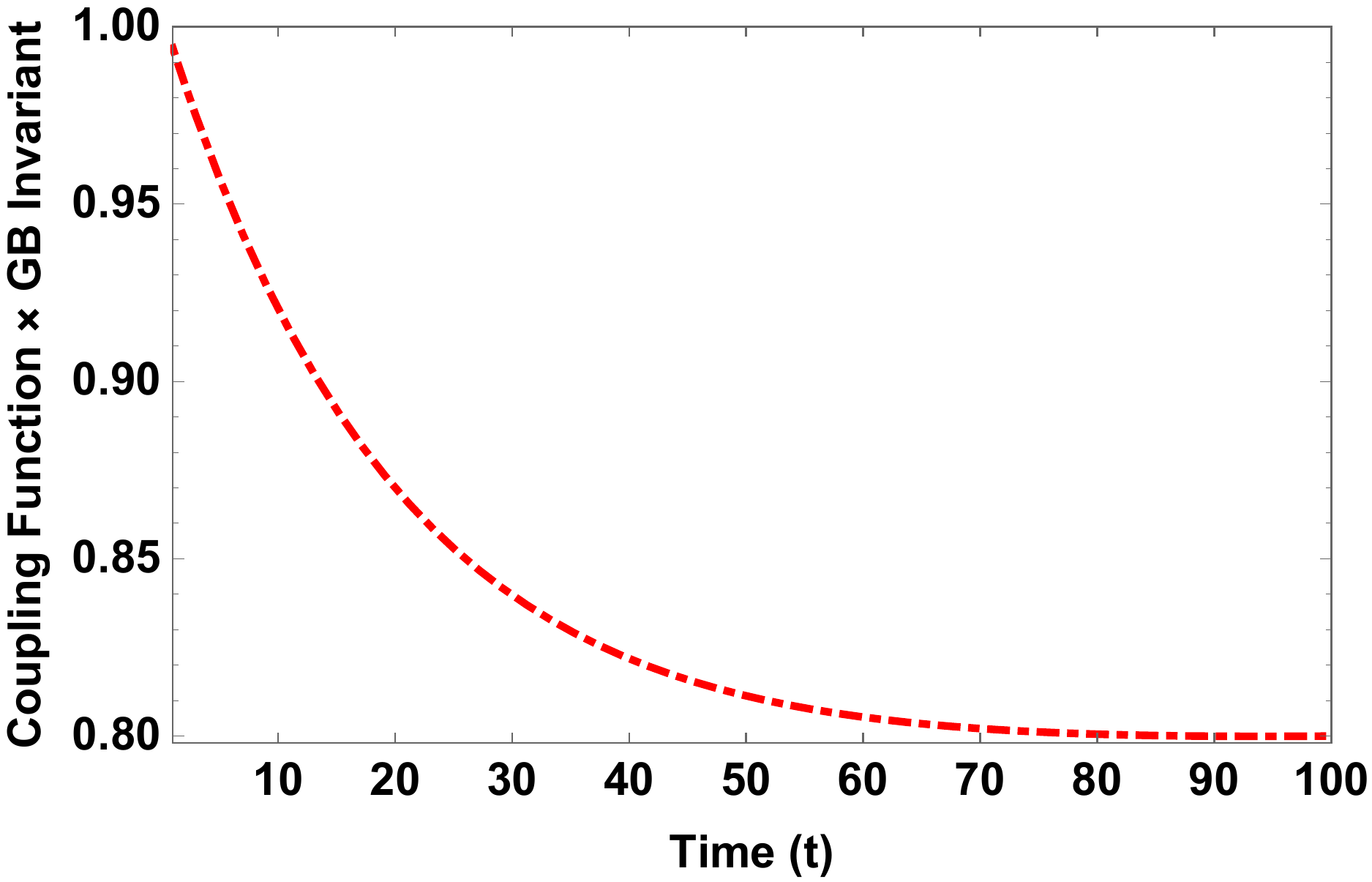}\\
\includegraphics[width=3in,height=2in]{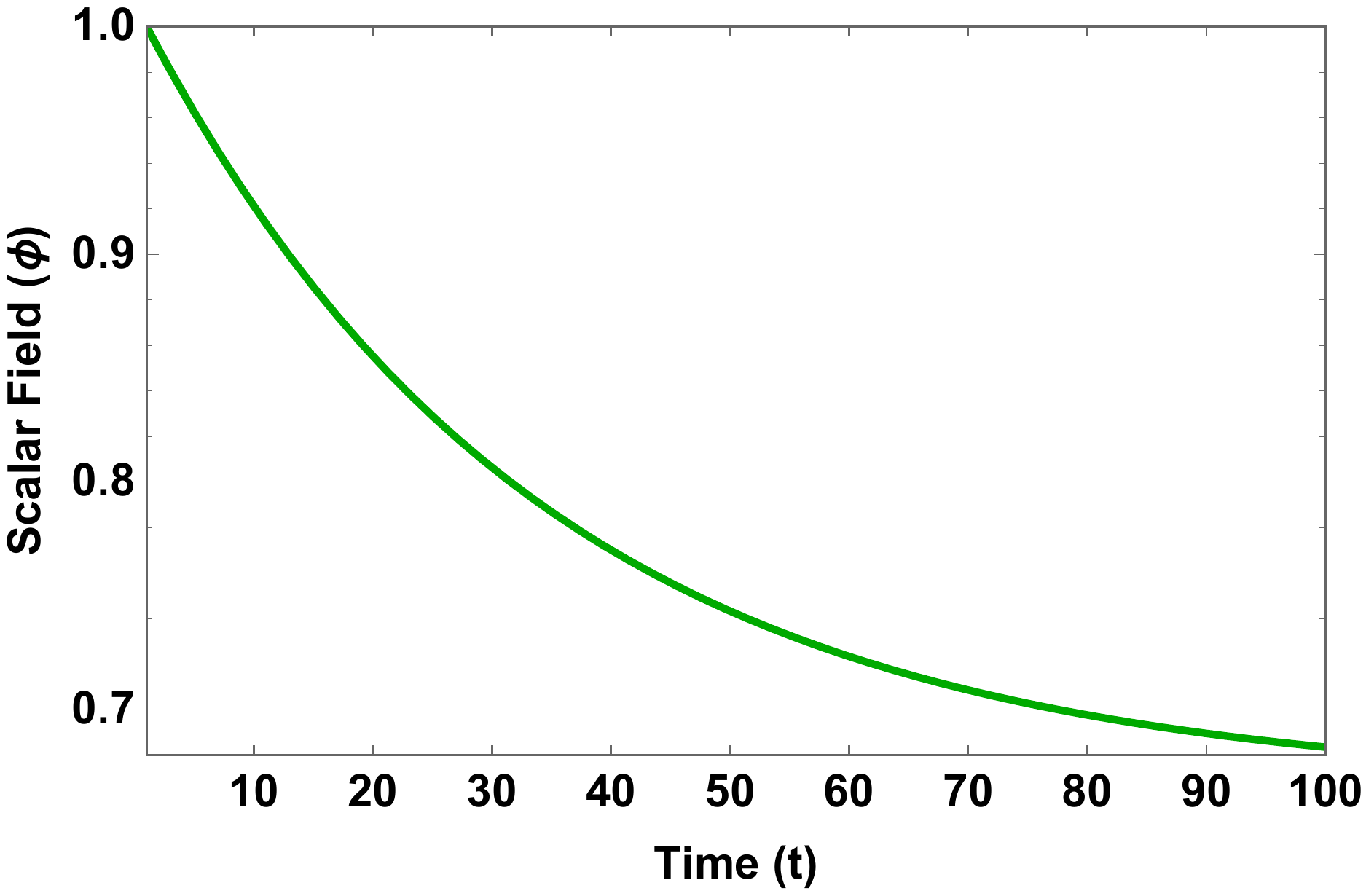}~~
\includegraphics[width=3in,height=2in]{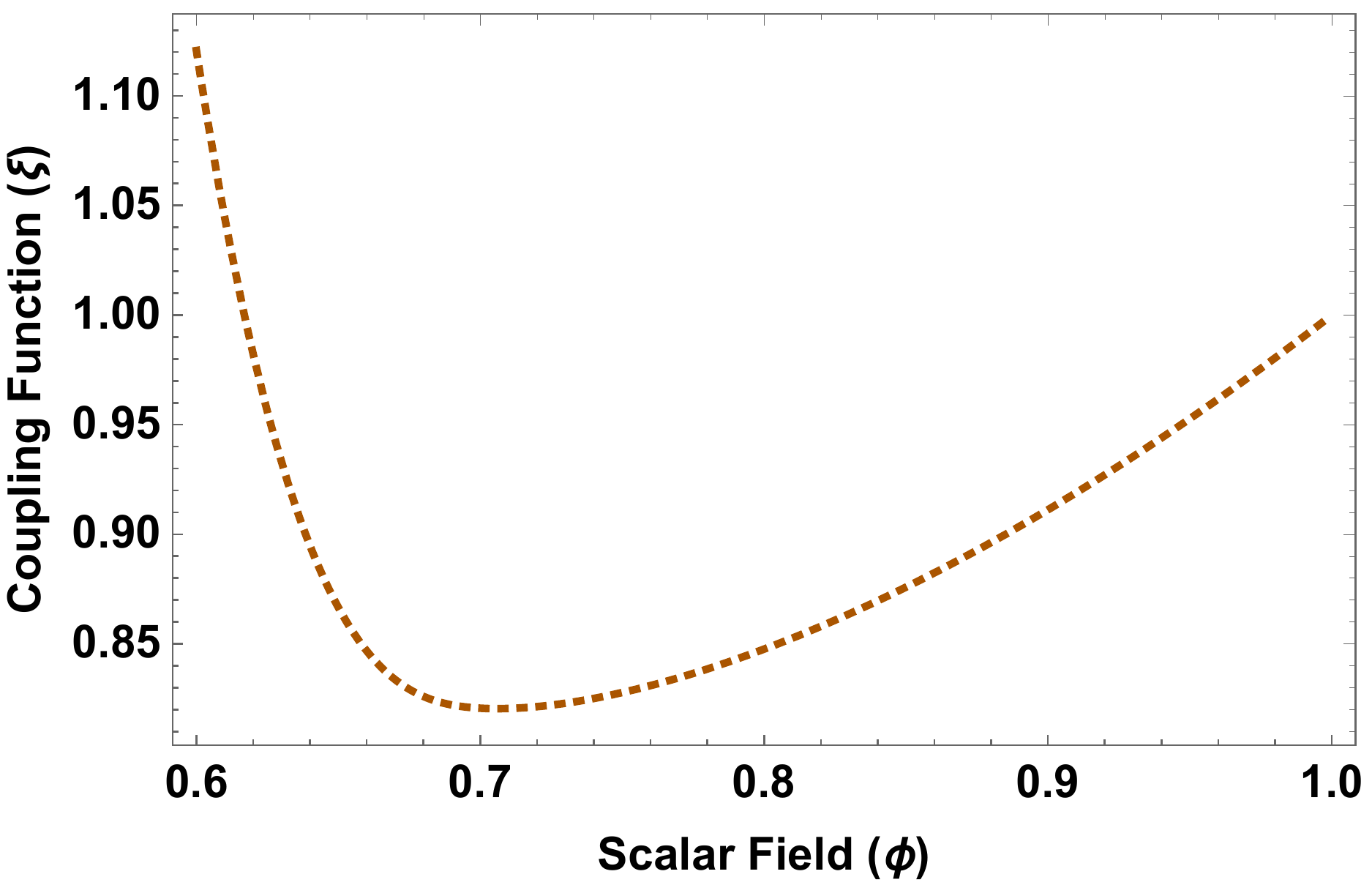}
\caption{The variation of the coupling function $\xi(\Phi)$, the contribution from Gauss-Bonnet invariant, i.e., $\xi(\Phi)\mathcal{G}$ and the scalar field $\Phi$ with time have been presented. Moreover, the variation of $\xi(\Phi)$ with $\Phi$ has also been depicted. All the plots are drawn by rescaling both the x and y axis to highlight the essential features. For example, the time coordinate has been rescaled by the definition $(t-t_{*})/t_{*}$ and hence has the above range. The coupling function $\xi(\Phi)$ is presented by the blue, dashed curve, which shows that it decreases with time, reaching a minima, while ultimately showing a feeble rise with time. On the other hand, the \GB invariant $\mathcal{G}$  decreases with time and as a consequence the term $\xi(\Phi)\mathcal{G}$ (depicted by the red, dot-dashed curved) also decreases with time and remained saturated at the final value. A similar behaviour is also seen in the time evolution for the scalar field $\Phi$ and is depicted by the green, thick curve. Finally, the variation of the coupling function $\xi(\Phi)$ with $\Phi$ has been presented by the dotted, orange curve. See text for more discussions.}
\label{figN_01}
\end{figure}

Given the Hubble parameter it is straightforward to obtain the differential equation determining the time evolution of the coupling function $\xi(\Phi)$ using \ref{Diff_Eq_xi}. The computation of individual coefficients of $\dot{\xi}$ and the $\xi$ independent term requires $\dot{H}$, which for the Hubble parameter presented in \ref{solution_of_H_2} with $\alpha \simeq 3/5$ becomes, $\dot{H}=(-3d/5)\{c-d(t-t_{*})\}^{-2/5}$. Therefore the differential equation satisfied by the potential $\xi(\Phi)$ becomes,
\begin{align}\label{diff_eq_xi}
2\ddot{\xi}+2\Big[5\left\{c-d(t-t_{*})\right\}^{3/5}-\frac{6d}{5\left\{c-d(t-t_{*})\right\}}\Big]\dot{\xi} +\Big[\frac{3d}{5\left\{c-d(t-t_{*})\right\}^{8/5}}-3\Big]=0~.
\end{align}
The above second order linear differential equation can be solved by evaluating the associated integrating factor, which in this scenario reads,
\begin{align}
\textrm{Integrating Factor}\equiv e^{P}&=\exp\Big\{\int dt \Big[5\left\{c-d(t-t_{*})\right\}^{3/5} -\frac{6d}{5\left\{c-d(t-t_{*})\right\}}\Big]\Big\}
\nonumber
\\
&=\exp\Big\{-\frac{25}{d}\frac{\left[c-d(t-t_{*})\right]^{8/5}}{8}
+\frac{6}{5}\ln \left[c-d(t-t_{*})\right]\Big\}
\nonumber
\\
&=\left\{c-d(t-t_{*})\right\}^{6/5}\exp \left[-\frac{25}{8d}\left\{c-d(t-t_{*})\right\}^{8/5}\right]~.
\end{align}
Therefore, multiplying the second order differential equation for the coupling function $\xi(t)$, presented in \ref{diff_eq_xi}, by the integrating factor it can be integrated once, resulting into,
\begin{align}\label{xi_diff_eq}
\dot{\xi}&=e^{-P(t)}\int _{t_{*}}^{t} dt \left[\frac{3}{2}-\frac{3d}{10\{c-d(t-t_{*})\}^{8/5}}\right]\left\{c-d(t-t_{*})\right\}^{6/5}\exp \left[-\frac{25}{8d}\left\{c-d(t-t_{*})\right\}^{8/5}\right]
\end{align}
which will result into incomplete Gamma functions. This in turn provides the expression for $\dot{\Phi}$ from \ref{gr_equation_temporal_part}. However, due to the complicated nature of the differential equations for $\xi(t)$ and $\Phi(t)$, as evident from \ref{xi_diff_eq}, it is not possible to obtain an analytic solution, unlike the case of constant Hubble parameter. Therefore, we have solved both the differential equations for $\xi(t)$ and $\Phi(t)$ using numerical techniques and have presented the results in \ref{figN_01}.

As evident from \ref{figN_01}, the scalar field decreases with time, which is expected, since at the beginning of the inflationary paradigm the scalar field was at the Planck scale, while as the inflation progresses the scalar field attains lower and lower values. An identical scenario also takes place for the coupling function $\xi(\Phi)$, which also shows a decreasing nature with time. Furthermore, if the \GB invariant is taken into account, the object $\xi(\Phi)\mathcal{G}$ starts decreasing with time. This is partly due to the decrease of $\xi(\Phi)$ but also due to the rapid fall of the \GB invariant with time, since the curvatures decreases rapidly with time as the inflation comes to an end. Finally, it is also clear from \ref{figN_01} that the coupling function $\xi(\Phi)$ initially decreases with the scalar field, which then starts increasing. This is because $\xi(\Phi)$ decreases with time at a slower pace than the scalar field itself. Therefore one can safely say that the \GB invariant alone is capable of driving the inflation. 
\section{Instability of the Model}\label{instab_GB}

It would have been really interesting if this becomes the end of the story. However unfortunately it turns out that despite having such intriguing features the above model faces a serious difficulty, namely stability against perturbations. In particular, for specific choices of the \GB coupling function it has been demonstrated that the tensor perturbations in the above spacetime grow rapidly \cite{Kawai:1998ab,Kawai:1999pw,Hikmawan:2015rze} and results into negative values for the sound speed. In particular, it was demonstrated in \cite{Kawai:1998ab,Kawai:1999pw} that cosmological solutions in models with only \GB coupling \emph{but} without a scalar potential are generically unstable if they are non-singular. Later on in \cite{Hikmawan:2015rze} it was demonstrated that the situation considered in \cite{Kanti:2015pda,Kanti:2015dra} is unstable as the sound speed becomes negative. In the present section we would like to present a general expression for the sound speed for arbitrary $\xi(\Phi)$ and explore the stability of tensor perturbations in absence of slow-roll approximations for the scalar field. On the other hand, in the context of scalar coupled Gauss-Bonnet theory, there are no growing scalar modes and the vector perturbations decrease as the universe expands \cite{Kawai:1997mf}. Thus to see the instability associated with tensor perturbations in a general context, we would like to analyse the sound speed associated with the evolution of tensor perturbations given the gravitational field equations. The 
tensor perturbations associated with a flat FRW background are given by:
\begin{eqnarray}
ds^2 = -dt^2 + a(t)^2\big(\delta_{ij}+h_{ij}\big)dx^idx^j
\label{new1}
\end{eqnarray}
where $h_{ij}(t,x^{k})$ stands for the tensor perturbation with transverse and traceless condition imposed on the same. Thus we have $\partial _{j}h^{ij}=0=h^i_i$. By substituting the perturbed metric presented in \ref{new1}, in the action of our model and expanding the action to quadratic order of the gravitational perturbation (in order to obtain 
field equations linear in $h_{ij}$) we obtain the ``perturbed'' action as follows \cite{Kawai:1997mf,Kawai:1998ab,Kawai:1999pw}:
\begin{eqnarray}
\mathcal{A}&=&\frac{1}{8}\int d^4x a^3 \bigg[\bigg(\dot{h}^{ij}\dot{h}_{ij} - \frac{1}{a^2}h_{ij,k}h^{ij,k} 
- (4\dot{H} + 6H^2 + \dot{\Phi}^2)h_{ij}h^{ij}\bigg)\nonumber\\
&+&4\ddot{\xi}\bigg(\frac{1}{a^2}h_{ij,k}h^{ij,k} + 2H^2h_{ij}h^{ij}\bigg)\nonumber\\
&-&4\dot{\xi}\bigg(-H\dot{h}^{ij}\dot{h}_{ij} + 4H(\dot{H}+H^2) h_{ij}h^{ij}\bigg)\bigg]
\label{new3}
 \end{eqnarray}
By using the background equations one can simplify the above action and it turns out to be,
\begin{eqnarray}
\mathcal{A}=\frac{1}{8}\int d^4x a^3 \bigg[(1-4H\dot{\xi}) \dot{h}^{ij}\dot{h}_{ij} - \frac{1}{a^2}(1-4\ddot{\xi})h_{ij,k}h^{ij,k}\bigg]
\label{new4}
\end{eqnarray}
At this stage it is advantageous to consider Fourier decomposition of the gravitational perturbation as: $h_{ij}(t,x^{k})=h_{ij}(t)\exp(ik^{l}x_{l})$ and hence the above expression of perturbed action (see \ref{new4}) leads to the following equation for time dependent part of tensor perturbation as,
\begin{eqnarray}
\ddot{h}_{ij} + \bigg(3H + \frac{1}{1-4H\dot{\xi}}\frac{d}{dt}(1-4H\dot{\xi})\bigg)\dot{h}_{ij} 
+\frac{k^2}{a^2}\bigg(\frac{1-4\ddot{\xi}}{1-4H\dot{\xi}}\bigg)h_{ij} = 0
\label{new5}
\end{eqnarray}
from where we can define the effective speed of sound as follows,
\begin{align}\label{sound_speed}
c_{\rm s}^{2}=\frac{1-4\ddot{\xi}}{1-4H\dot{\xi}}~,
\end{align}
where $\dot{\xi}=(\partial \xi/\partial \Phi)\dot{\Phi}$. The expression for $\ddot{\xi}$ can also be derived from \ref{Diff_Eq_xi} and can be used to obtain,
\begin{align}
1-4\ddot{\xi}&=1-4\left[\left(5H+2\frac{\dot{H}}{H}\right)\dot{\xi}-\left(\frac{\dot{H}}{2H^{2}}+\frac{3}{2}\right) \right]
=\left(2\epsilon -5\right)\left(1-4H\dot{\xi}\right)~,
\end{align}
where, $\epsilon$ is the slow-roll parameter $-\dot{H}/H^{2}$. The above expression when substituted in \ref{sound_speed} for sound speed yields, 
\begin{align}
c_{\rm s}^{2}=2\epsilon -5
\end{align}
Therefore throughout the inflationary epoch, we have $\epsilon \ll 1$ and hence $c_{\rm s}^{2}$ is negative. Note that the existence of instability in tensor perturbations has been inferred earlier for specific choices of the \GB coupling function, while the above derivation is general and holds for \emph{all} possible choices of $\xi(\Phi)$ and \emph{without} any slow-roll approximation. Thus irrespective of the choice of the \GB coupling function $\xi(\Phi)$ there is an instability in the tensor perturbation. As a consequence the fluctuations in the tensor modes will grow rapidly and hence the above model without a self-interacting potential for the inflaton field can not lead to a viable inflationary scenario. Thus it is necessary to include a self-interacting term in the Lagrangian in order to explain the behaviour of the perturbations in a consistent manner. For completeness we would like to present the corresponding expression for sound speed in presence of self-interacting potential. Since the scalar and vector perturbations were not problematic, we will consider tensor perturbations only in our analysis. Regarding the same, if we go through the same calculational steps as discussed in the earlier section, we finally end up with the following expression of ``effective speed of sound'' in presence of self-interacting scalar potential $V(\Phi)$ as,
\begin{eqnarray}
c_s^2 = \big(2\epsilon - 5\big) + \frac{2V(\Phi)}{H^{2}\left(1 - 4H\dot{\xi}\right)}
\label{new7}
\end{eqnarray}
During inflationary era, $\epsilon$ is less than unity and hence $2\epsilon-5$ remains negative, while due to the presence of the potential term $V(\Phi)$, $c_s^2$ may become positive and thereby leads to a stable inflationary scenario, unlike the situation of without the self-interacting potential.
\section{Inflation with a self-interacting potential}\label{Inf_GB_Pot}

We have just described the instability of the tensor perturbation in absence of a self-interacting potential for the scalar field, this being a strong motivation towards introduction of such a self-interacting potential, even though the scalar coupled \GB term alone can provide a consistent inflationary scenario (keeping aside the perturbations). Thus in this section we will explore the possible solutions of the field equations consistent with inflationary paradigm in presence of such a self-interacting potential. This will result into modifications of the gravitational field equations, which in turn will modify \ref{gr_equation_temporal_part} and \ref{scalar_field_equation} respectively, while \ref{gr_equation_spatial_part} will remain unchanged. In particular the right hand side of \ref{gr_equation_temporal_part} will get modified by the introduction of $8\pi G~V(\Phi)$ term, while the left hand side of \ref{scalar_field_equation} will inherit an additional $\partial V/\partial \Phi$ term, such that 
\begin{align}
3H^{2}-12H^{3}\dot{\xi}&=8\pi G\left(\frac{1}{2}\dot{\Phi}^{2}+V(\Phi)\right);
\label{gr_equation_temporal_part_N}
\\
2\dot{H}-4H^2\Big[\ddot{\xi}-H\dot{\xi}&+2\frac{\dot{H}}{H}\dot{\xi}\Big]=-8\pi G~\dot{\Phi}^2;
\label{gr_equation_spatial_part_N}
\\
\ddot{\Phi}+3H\dot{\Phi}+\frac{12H^{2}}{8\pi G}\Big(H^{2}&+\dot{H}\Big)\frac{\partial \xi}{\partial \Phi}+\frac{\partial V}{\partial \Phi}=0~.
\label{scalar_field_equation_N}
\end{align}

Given these modifications we are now in a position to study effect of both these terms on the inflationary epoch. Alike the previous scenario with the \GB term alone, in the present context as well the inflationary paradigm and slow-roll approximation for the scalar field are \emph{incompatible} with each other as we will demonstrate below. In the slow-roll approximation we neglect $\ddot{\Phi}$ and $\dot{\Phi}^{2}$ terms in comparison with $\Phi$ and hence the field equation as in \ref{gr_equation_temporal_part_N} yields,
\begin{align}
\dot{\Phi}=\frac{3H^{2}-8\pi G V(\Phi)}{12 H^{3}(\partial \xi/\partial \Phi)}~.
\end{align}
The above expression for $\dot{\Phi}$ must be contrasted with the corresponding situation in absence of the \GB term, where the same equation would result into $H^{2}\sim V(\Phi)$ and $\dot{\Phi}\simeq 0$. Thus the presence of the \GB coupling essentially makes the time derivative of the scalar field, namely the term $\dot{\Phi}$ to be non-zero and finite throughout the inflationary scenario. On the other hand, $\dot{H}$ can be obtained from \ref{gr_equation_spatial_part_N}, such that the slow-roll parameter becomes, 
\begin{align}
\epsilon \equiv-\frac{\dot{H}}{H^{2}}\simeq \frac{2H(\partial \xi/\partial \Phi)\dot{\Phi}}{1-4H\dot{\Phi}(\partial \xi/\partial \Phi)}
=\frac{3H^{2}}{16\pi G V}-\frac{1}{2}~.
\end{align}
Thus if we neglect the \GB term then of course this is a very small quantity and the normal inflationary paradigm would follow. But in presence of the \GB term the above slow-roll parameter is always $\sim \mathcal{O}(1)$ and hence it is not possible to have accelerated expansion of the universe while respecting slow-roll approximation. Thus one must abandon the slow-roll approximation if the non-trivial effects of the \GB term in the early universe cosmology is asked for. This suggests to take an identical route as in the previous scenario. However due to the complicated nature of the field equations, unlike the previous situation here we will not employ the reconstruction scheme, rather should provide viable choices for the potential $V(\Phi)$ as well as the coupling function $\xi(\Phi)$ for which analytical solutions can be obtained. We would again like to emphasize that we are \emph{not} neglecting the \GB term while considering inflationary paradigm, rather we are keeping both the self-interacting potential and the \GB term to have an initial accelerated expansion of the universe as well as a final deceleration signifying end of the inflationary epoch. 
\subsection{Accelerated expansion with a quadratic potential}\label{Tan_First}

As a first choice it is convenient to consider a quadratic potential for the scalar field, i.e., the potential function $V(\Phi)$ involves a constant contribution and a quadratic part proportional to $\Phi ^{2}$. A similar form for the coupling function $\xi(\Phi)$ is also suggestive. However the field equations involves derivative of $\xi(\Phi)$ and hence the constant term in $\xi(\Phi)$ plays no role. This implies the following form of the scalar field potential and the coupling function,
\begin{align}
 V_{1}(\Phi)&=V_{0}^{(1)}+V_{1}^{(1)}\Phi ^{2};
 \label{potential_01}
 \\
 \xi_{1}(\Phi)&=\xi _{0}^{(1)}\Phi ^{2}~,
 \label{coupling_01}
\end{align}
where the subscript `1' denotes that the above corresponds to the first set of solutions. Furthermore, $V_{0}^{(1)}$, $V_{1}^{(1)}$ and $\xi _{0}^{(1)}$ stands for arbitrary parameters in the theory, which needs to be determined later. Substituting the above form of the potential function $V_{1}(\Phi)$ and $\xi_{1}(\Phi)$ into the field equations, one easily obtains the following solutions of the scalar field and the Hubble parameter as,
\begin{align}
H(t)&=H_{0}\equiv \sqrt{\frac{8\pi G~V_{0}^{(1)}}{3}}=\textrm{constant};
\label{Hubble}
\\
\Phi(t)&=\Phi_{0}\exp{(-\lambda t)}~.
 \label{scalar_solution_1}
\end{align}
Here, the unknown parameters namely $\lambda$ and $V_{1}^{(1)}$ can be obtained in terms of the constant Hubble parameter $H_{0}$ as well as $\xi_{0}^{(1)}$ as, 
\begin{align}\label{parameter_01}
\lambda =\frac{8H_{0}^{3}\xi _{0}^{(1)}}{8\pi G-16H_{0}^{2}\xi _{0}^{(1)}};\qquad V_{1}^{(1)}=\frac{24H_{0}^{3}}{8\pi G}\lambda \xi _{0}^{(1)}-\frac{\lambda ^{2}}{2}~,
\end{align}
while the parameter $\Phi_{0}$ remains undetermined.  

This solution can also be derived using the reconstruction scheme advocated in \cite{Cognola:2006sp} in the context of Einstein-scalar-Gauss-Bonnet gravity. This is achieved by introducing an additional quantity $W(t)$, defined as 
\begin{align}\label{W_function}
W(t)=\int ^{t}dt'\frac{1}{a(t')}\left[\frac{\dot{H}(t')}{4\pi G}+\dot{\Phi}^{2}(t')\right]~,
\end{align}
in terms of which the scalar potential as well as the coupling function gets determined. In this particular case, with the choices of the Hubble parameter $H(t)$ and the scalar field $\Phi(t)$ as in \ref{Hubble} and \ref{scalar_solution_1}, the above function becomes, $W(t)=A-B\exp\{(-H_{0}+2\lambda) t\}$, where $A$ is an integration constant and $B$ is dependent on $H_{0}$, $\lambda$ and $\Phi_{0}$. Following \cite{Cognola:2006sp}, one can immediately verify that, the associated scalar potential and the coupling function has the desired behaviour, i.e., their behaviours are identical to those presented in \ref{potential_01} and \ref{coupling_01}, provided $A$ vanishes. Thus the results presented in this section are indeed consistent with those presented in \cite{Cognola:2006sp}. 

At this stage it would be worthwhile to briefly mention about the attractor nature of the solution presented above. This essentially implies that even under small perturbations the solutions will ultimately converge to the ones given above. In other words, the perturbations must die down as time progresses. As demonstrated in \cite{Guo:2010jr}, by rewriting the gravitational field equations, any perturbations around de-Sitter background decays with time with additional corrections depending on $\epsilon$. Thus as long as $\epsilon$ is smaller we will have the perturbations decaying exponentially with time, resulting into the stability of the de Sitter solution. Thus even in the context of \GB coupled scalar field the de Sitter solution remains an attractor.

As evident, constant value for the Hubble parameter ensures that the scale factor scales exponentially with time, i.e., $a(t)=\exp(H_{0}t)$. Thus the solution corresponds to accelerating phase of the universe. Furthermore it is straightforward to determine the time evolution of the self-interacting potential $V_{1}(\Phi)$ as well as the coupling function $\xi_{1}(\Phi)$ using the time evolution of the scalar field. This ensures that $V(\Phi)$ has a constant piece and the rest of the part decays exponentially with time, while $\xi(\Phi)$ also decays exponentially. Thus at later stages of inflation these potentials must be replaced with some other scalar potentials, allowing for decelerated expansion of the universe, which we consider in the subsequent section.
\subsection{Power law expansion and deceleration}

In this section we will discuss another set of solutions for the scalar field and the scale factor, given some appropriate form for the scalar potential as well as the coupling function. We assume that the potential is an exponentially decaying function of the scalar field, while the coupling function is an exponentially growing one. The growing behaviour is necessary since we would like to keep the \GB term relevant even at the end stages of inflation. (Note that the \GB term alone should have negligible contribution at the end of inflation as the curvatures has become quite small.) Thus for our purpose we consider a different form of the scalar field potential and the coupling function,
\begin{align}
V_2(\Phi)&=V_{0}^{(2)}\exp{[-2\Phi(t)/\Phi_{0}]};
\label{eq_potential_02}
\\
\xi_2(\Phi)&=\xi_{0}^{(2)}\exp{[2\Phi(t)/\Phi_{0}]}~,
\label{eq_coupling_02}
\end{align}
where the subscript `2' is just to remind us that this corresponds to the second set of solutions. In the above expression $V_{0}^{(2)}$, $\Phi_{0}$ and $\xi_{0}^{(2)}$ are the model parameters. It can be easily verified that the field equations for gravity plus scalar field is satisfied provided the time dependence of the scale factor and the scalar field corresponds to 
\begin{align}\label{sol_phi_h_02}
\Phi(t)=\Phi_{0}\ln{(t/t_{0})};\qquad H(t)=n/t~,
\end{align}
where $n<1$. One can easily check that $\dot{H}+H^{2}$ for this particular case is negative and thus corresponds to the decelerating scenario at the end of the inflation. Since it is normally believed that the end of inflation results into a radiation dominated universe, it is legitimate to assume $n=1/2$. However for the moment we will keep $n$ arbitrary. The field equations also result into several constraints connecting the free parameters present in the model. In particular, the parameter $\xi _{0}^{(2)}$ and $V_{0}^{(2)}$ gets determined in terms of the other free parameters as,
\begin{align}\label{parameters_02}
\frac{\xi _{0}}{t_{0}^{2}}=\frac{8\pi G}{24n^{3}(n-1)}\left[\left(1-3n\right)\Phi_{0}^{2}t_{0}+2V_{0}t_{0}^{2}\right];\qquad
V_{0}t_{0}^{3}=\frac{\left(n-1\right)}{2}\left[\frac{3n^{2}}{8\pi G}-\frac{\Phi_{0}^{2}t_{0}^{2}}{2(n-1)}\left(1-5n\right)\right]~.
\end{align}
Finally plugging the solution for the time evolution of the scalar field into the expressions for the self-interacting potential as well as coupling function one gets both of them as a function of time:
\begin{align}
V_{2}\left[\Phi(t)\right] = V_{0}^{(2)}\left(\frac{t_{0}^{2}}{t^{2}}\right);\qquad \xi_{2}\left[\Phi(t)\right]=\xi_{0}^{(2)}\left(\frac{t^{2}}{t_{0}^{2}}\right)~.
\end{align}
Thus as in the previous scenario here also the scalar field potential decays with time but as a power law, while the interaction potential depicts a growth with time. This behaviour of the potential as well as that of the coupling function can again be derived using the reconstruction scheme advocated in \cite{Cognola:2006sp}. For example, with the Hubble parameter and the scalar field presented in \ref{sol_phi_h_02}, following \ref{W_function}, the function $W(t)$ can be determined to be, $A+Bt^{-n-1}$. For $A=0$, this reproduces the  structure of the scalar potential and the coupling function as in \ref{eq_potential_02} and \ref{eq_coupling_02}. This once again demonstrates the validity of these results even in the reconstruction scheme.
\subsection{Estimation of parameters associated with the inflationary scenario}

Having described the two situations, one depicting accelerated expansion of the universe at the early stages of inflation and the other providing a decelerating phase marking the exit from inflationary paradigm, we concentrate on estimation of various parameters in the model. The inflationary paradigm comes into existence at very early stages of the universe and it lasted from $t_{\rm in}\sim 10^{-11}~\textrm{GeV}^{-1}$ to $t_{\rm end}\sim 6\times10^{-8}~\textrm{GeV}^{-1}$. Thus we assume that the potential $V_{1}(\Phi)$ existed for an initial phase of the inflationary epoch which we choose to be in the range $10^{-11}~\textrm{GeV}^{-1}<t<10^{-8}~\textrm{GeV}^{-1}$, while the other potential $V_{2}(\Phi)$ appeared in the end stages of the inflationary scenario and was effective for $t>6\times 10^{-8}~\textrm{GeV}^{-1}$. During the regime $10^{-8}~\textrm{GeV}^{-1}<t<6\times 10^{-8}~\textrm{GeV}^{-1}$, there must be an intermediate potential interpolating between these two regimes, which we will determine 
later using numerical techniques. Along identical lines the coupling potential $\xi(\Phi)$ also has two different behaviour in the two distinct regimes. We will have $\xi(\Phi)=\xi_{1}(\Phi)$ for $10^{-11}~\textrm{GeV}^{-1}<t<10^{-8}~\textrm{GeV}^{-1}$, while the coupling function becomes, $\xi(\Phi)=\xi_{2}(\Phi)$ for $t>6\times 10^{-8}~\textrm{GeV}^{-1}$. In the intermediate region we will numerically construct an interpolating coupling function that matches with both $\xi_{1}(\Phi)$ and $\xi_{2}(\Phi)$ appropriately at both ends.

The above process of interpolation requires appropriate choices for the values of the free parameters present in our model. As far as the first situation is considered, the relevant parameters are the Hubble parameter $H_{0}$ and the decaying parameter $\lambda$ in the solution of the scalar field (see \ref{scalar_solution_1} for a detailed description), both having mass dimension one. The choice of these parameters are also connected with the observational viability of this model and hence it must have number of e-foldings $\sim 60$. Since the number of e-foldings correspond to integration of Hubble parameter over the entire duration of inflation, it follows that $H_{0}\simeq 6\times 10^{9}~\textrm{GeV}$. 

Using the scalar field solution presented in \ref{scalar_solution_1}, one can immediately verify that the energy density ($\rho$) of the scalar field $\Phi$ varies as $\rho \sim \exp{(-2\lambda t)}$ with time. Since, alike the scale factor, the energy density of the scalar field as well is supposed to decrease by a factor of $\sim 10^{15}$ starting from the beginning of the inflationary epoch to its end, it is legitimate to take $\lambda \sim 10^{9}~\textrm{GeV}$, of the same order as the Hubble parameter. A better estimate for the energy density of the scalar field would require its equation of state parameter, which can be used to relate $\lambda$ to the associated Hubble parameter $H_{0}$. Since in this scenario the equation of state parameter can not be defined in a simple manner, it must be obtained by numerical evolution of the Einstein's equations in the present context. However, as exact estimations of various parameters are not of much relevance to the present work, we will content ourselves with the above estimate of the parameter $\lambda$. Similarly, using \ref{parameter_01}, we immediately obtain both $\xi_{0}^{(1)}$ and $V_{1}^{(1)}$ in terms of $H_{0}$ and $\lambda$, leading to possible numerical estimates of both these parameters.

Returning to the post inflationary scenario we concentrate on the second set of solution given by the the potential $V_{2}(\Phi)$ and $\xi_{2}(\Phi)$ respectively, presented in \ref{eq_potential_02} and \ref{eq_coupling_02}. As evident we can choose the initial time instant to be located at $t_{0}\sim 10^{-8}\textrm{GeV}$ and hence the parameter $V_{0}^{(2)}$ gets determined from \ref{parameters_02} as $V_{0}^{(2)}t_{0}^{3}\simeq (1/8\pi G)$. The rest of the parameters can also be accordingly determined. As a consequence we can interpolate both the potential and the coupling function in the intermediate region.
\subsection{Numerical solutions in the interpolating region}

Given the structure of the potential as well as the coupling function in the initial and final stages of inflation, we would like to provide a complete picture by interpolating between these regions. Due to complicated nature of the equations governing the evolution of the scalar field and the scale factor in a general context, we will determine the interpolating function using numerical techniques and shall illustrate the same. Let us briefly point out the methods one may use in order to generate such interpolating solutions. In the intermediate region, one approximates the behaviour of the physical quantity of interest (e.g., the coupling function $\xi(\Phi)$ or the scalar field $\Phi$) by a polynomial function of time, with degree of the polynomial kept arbitrary. Then in the initial epoch one uses the analytic behaviour of the desired physical quantity (e.g., the scalar potential) to generate numerical estimates of the respective quantity at various time instants till the description is reliable. Similar numerical estimations are being made at the end stage of inflation as well. With these sets of initial and final data and the polynomial function one can use any standard interpolation package (e.g., MATHEMATICA) to end up getting the desired plots. The structure of the plot of course depends on the degree of the polynomial and desired accuracy level. All the plots in this paper are for a accuracy level of $\mathcal{O}(10^{-7})$. This procedure is repeated for all the remaining variables of interest as well. However the details of the interpolation of the curve connecting the initial instants of inflation to the end stage of inflationary scenario is an artefact of the procedure followed and admits possible variations depending on the process of interpolation by numerical techniques. Since our aim is essentially to demonstrate that interpolating functions satisfying the initial and the final stages of inflation as modeled here indeed exists, such indeterminacy in determining the interpolating function would not affect the results presented here. Finally when variation of all the variables with time has been obtained, one can use an analogue of the parametric plot to illustrate variation of the scalar potential and scalar coupling function with scalar field itself. As a further check of the results, we have verified that the plots generated by interpolation in the vanishing potential limit exactly matches with those presented in \ref{Inf_self_int}. Thus having explained the details of the interpolating procedure, we now turn to the corresponding implications and present the variations of all the relevant parameters with time.

\begin{figure}[H]
\begin{center}
\centering
\includegraphics[scale=1.2]{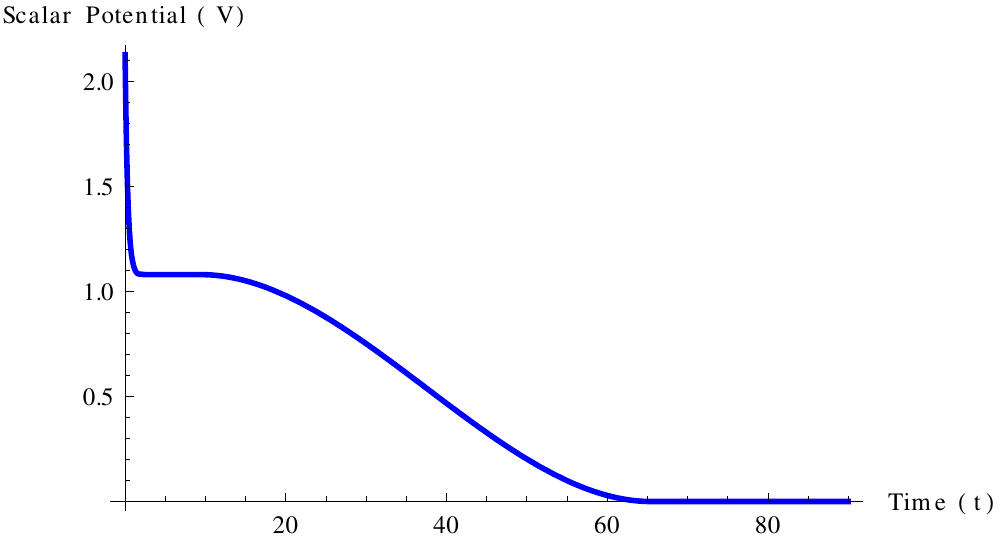}
\caption{The self-interacting scalar Potential $V(\Phi)$ is being plotted against time $t$ for the complete duration of inflation. The initial and final portions are determined analytically, while the intermediate region is obtained by interpolation. The curve explicitly shows the decreasing behaviour of the scalar potential with time. }
\label{fig_01}
\end{center}
\end{figure}

In particular, taking the Planck mass to be $M_{\rm pl}=10^{19}$ GeV and the expressions for potential in the early and late stages of inflation, we interpolate the potential function for $10^{-11} < t < 6\times10^{-8}~\textrm{GeV}^{-1}$, which has been presented in \ref{fig_01}. Note that the axes in \ref{fig_01} are rescaled according to convenience, namely x-axis corresponds to a ``rescaled" time coordinate obtained as $\sim 10^{9}t$ which is in $\textrm{GeV}^{-1}$ unit, while the y axis corresponds to ``rescaled" potential, which is in $\textrm{GeV}^{4}$ unit. It is evident that the potential function is smooth everywhere and decays with time.

\begin{figure}[H]
\begin{center}
\centering
\includegraphics[scale=1.2]{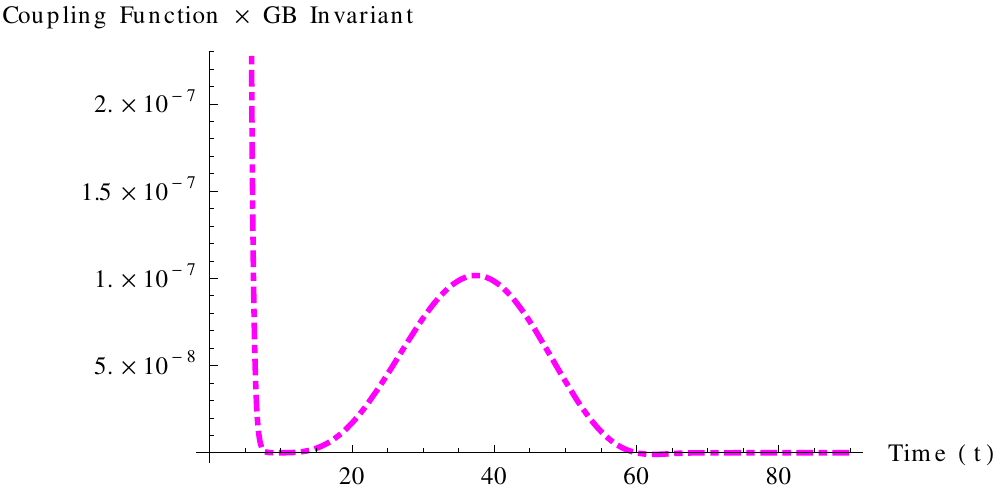}
\caption{The coupling function $\xi$ multiplied with \GB invariant $\mathcal{G}$ is being plotted against time $t$. The figure shows an initial decrease, with a subsequent increase in value, which finally decreases and gets saturated. Therefore in the intermediate region during the inflationary epoch, at some stage (around $t\sim 4\times 10^{-8} ~\textrm{GeV}^{-1}$) the additional contribution due to the \GB term attains a maximum value.}
\label{fig_02}
\end{center}
\end{figure}

Similarly substituting the values of various parameters presented into \ref{coupling_01} and \ref{eq_coupling_02}, one gets the coupling $\xi(\Phi)$ within the two time scales, $10^{-11}<t<10^{-8}~\textrm{GeV}^{-1}$ as well as for $\xi(\Phi)$ with $t > 6\times10^{-8}~\textrm{GeV}^{-1}$ respectively. Using the above two expressions, the time variation of the coupling function for the intermediate region can also be determined by interpolation. However rather than the coupling function, the combination $\xi(\Phi)\mathcal{G}$, where $\mathcal{G}$ is the Gauss-Bonnet invariant is of more importance and has been presented in \ref{fig_02}, where the x axis correspond to ``rescaled" time. As evident from \ref{fig_02} there exist an intermediate region where the effect of the coupling function times the Gauss-Bonnet invariant attains a maximum value. Thus during the inflationary epoch it is not at all justified to ignore the effect of the \GB term. On the other hand, as the universe exits from the inflationary epoch, the combination attains a fairly constant value and thus one may use it in the context of quintessential inflation. By using these forms of the scalar field potential and the coupling function, we are next going to solve the field equations for the Hubble parameter (or, equivalently the scale factor) as well as the scalar field numerically to understand their behaviour. 

\begin{figure}[H]
\begin{center}
\centering
\includegraphics[scale=0.95]{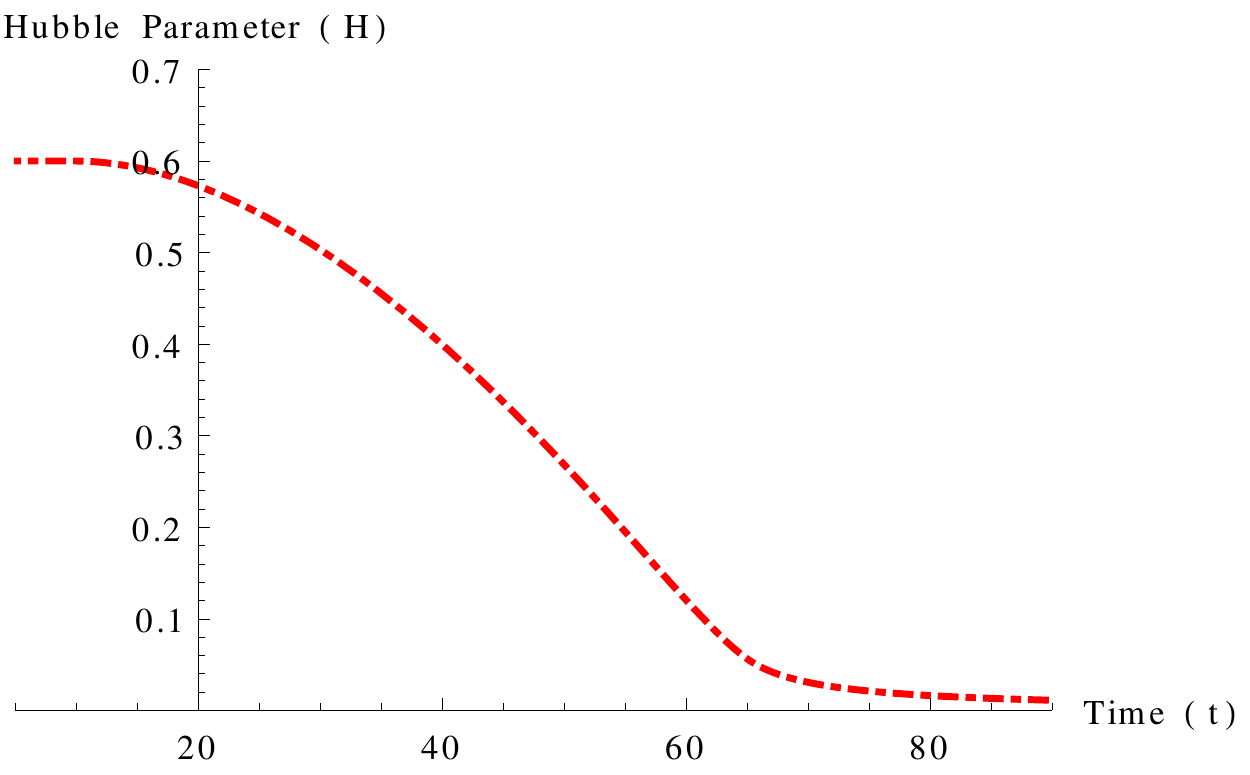}
\caption{Numerical solution of the Hubble parameter $H$ is being presented with time $t$. As evident at the onset of inflation, the Hubble parameter was fixed at a constant value, signifying initial exponential expansion of the universe, which then give way to final power law expansion. The behaviour of the Hubble parameter in the intermediate regime has been obtained by appropriate interpolation of the initial and final phases.}
\label{fig_03}
\end{center}
\end{figure}

Given the gravitational field equations involving only first order time derivatives of the Hubble parameter $H(t)$, a numerical solution of the same requires one boundary condition. Choosing the initial value of the Hubble parameter $H(t)$ as the inverse of the duration of the inflationary epoch i.e., $H(0)\sim 0.6\times10^{9}~\textrm{GeV}$, we obtain the required solution as depicted in \ref{fig_03}. As in the earlier plots, in \ref{fig_03} as well the x and y axes are rescaled such that the ``rescaled" Hubble parameter $\sim10^{-9}H$ in $\textrm{GeV}$ unit. The figure explicitly demonstrates that the Hubble parameter at the initial stages remained almost constant, signifying a very small value for the parameter $\epsilon(t)$, while at the later stages the Hubble parameter decreases with time and finally results into deceleration signifying an exit from inflationary paradigm. Thus we can safely argue that the numerical solutions obtained above indeed matches with the analytic one both at the beginning and at the end of the intermediate region.

\begin{figure}[H]
\begin{center}
\centering
\includegraphics[scale=1.2]{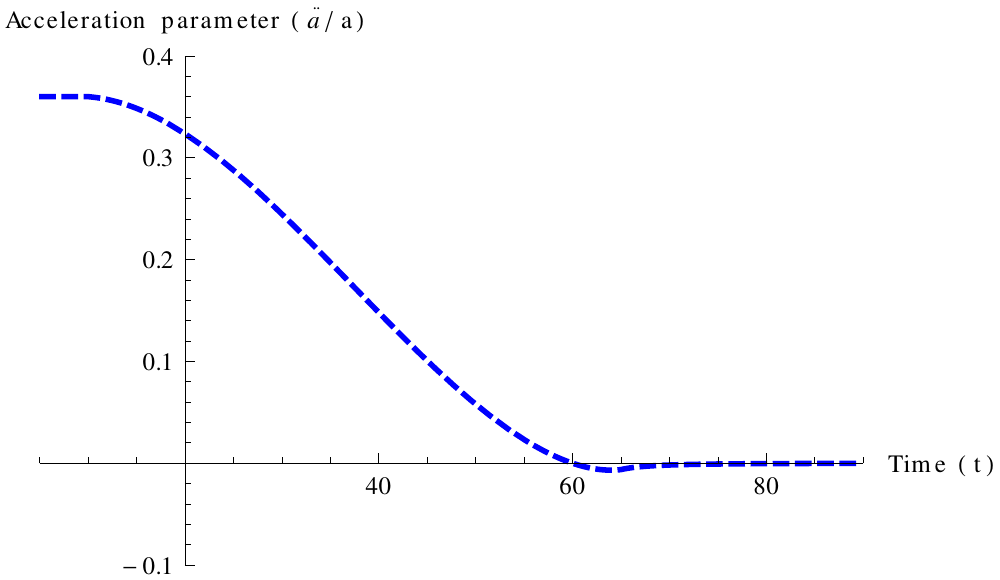}
\caption{The above figure presents the variation of the acceleration parameter $\ddot{a}/a$ with time. As evident in the initial stages of inflation, the acceleration was almost constant, while the acceleration decreases as time passes by and finally it turns negative around $t\sim 6\times 10^{-8}~ \textrm{GeV}^{-1}$. This presents the exit from inflation.}
\label{fig_04}
\end{center}
\end{figure}

\begin{figure}[H]
\begin{center}
\centering
\includegraphics[scale=1.2]{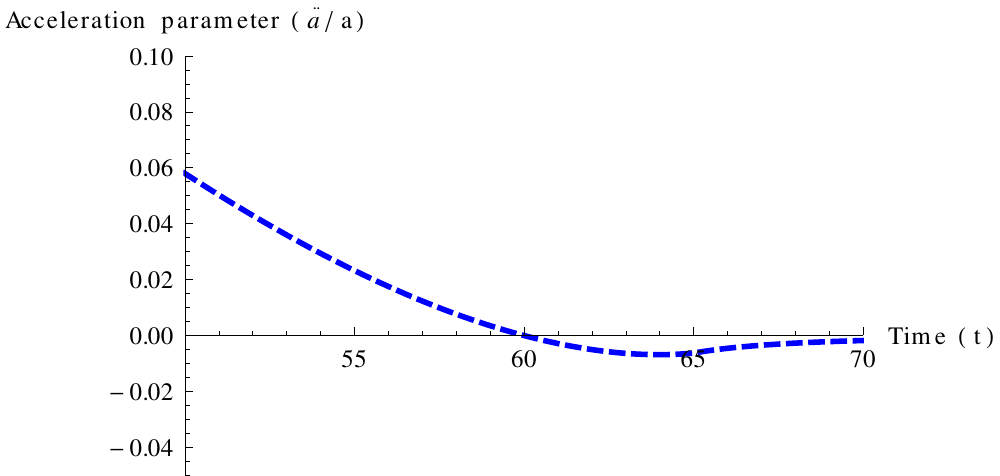}
\caption{A magnified plot depicting $\ddot{a}/a$ turning negative near the end of the inflationary paradigm, where a transition from acceleration to deceleration takes place. In this context the exit from inflation happened roughly when $t\sim 6\times 10^{-8}~\textrm{GeV}^{-1}$.}
\label{fig_05}
\end{center}
\end{figure}

The above numerical solution of the Hubble parameter can be immediately integrated providing the evolution of scale factor $a(t)$ with respect to time. However in the context of inflation it is more convenient to depict the solution for $\ddot{a}/a$, the acceleration parameter of the universe, which has been presented in \ref{fig_04}. Here the y-axis of \ref{fig_04} corresponds to $\ddot{a}(t)/a(t)$ associated with the ``rescaled" Hubble parameter. From the above figure, one can easily conclude that the inflation ends near about $\bar{t} \sim 6\times 10^{-8}~\textrm{GeV}^{-1}$ or, equivalently $t \simeq 6\times10^{-32}~\textrm{sec}$, after which $\ddot{a}/a$ becomes negative. To get a better view of what is happening near the end of the inflationary epoch, we provide in \ref{fig_05} a zoomed-in version of \ref{fig_04} near $t \sim 6\times 10^{-8}~\textrm{GeV}^{-1}$.

\begin{figure}[H]
\begin{center}
\centering
\includegraphics[scale=0.6]{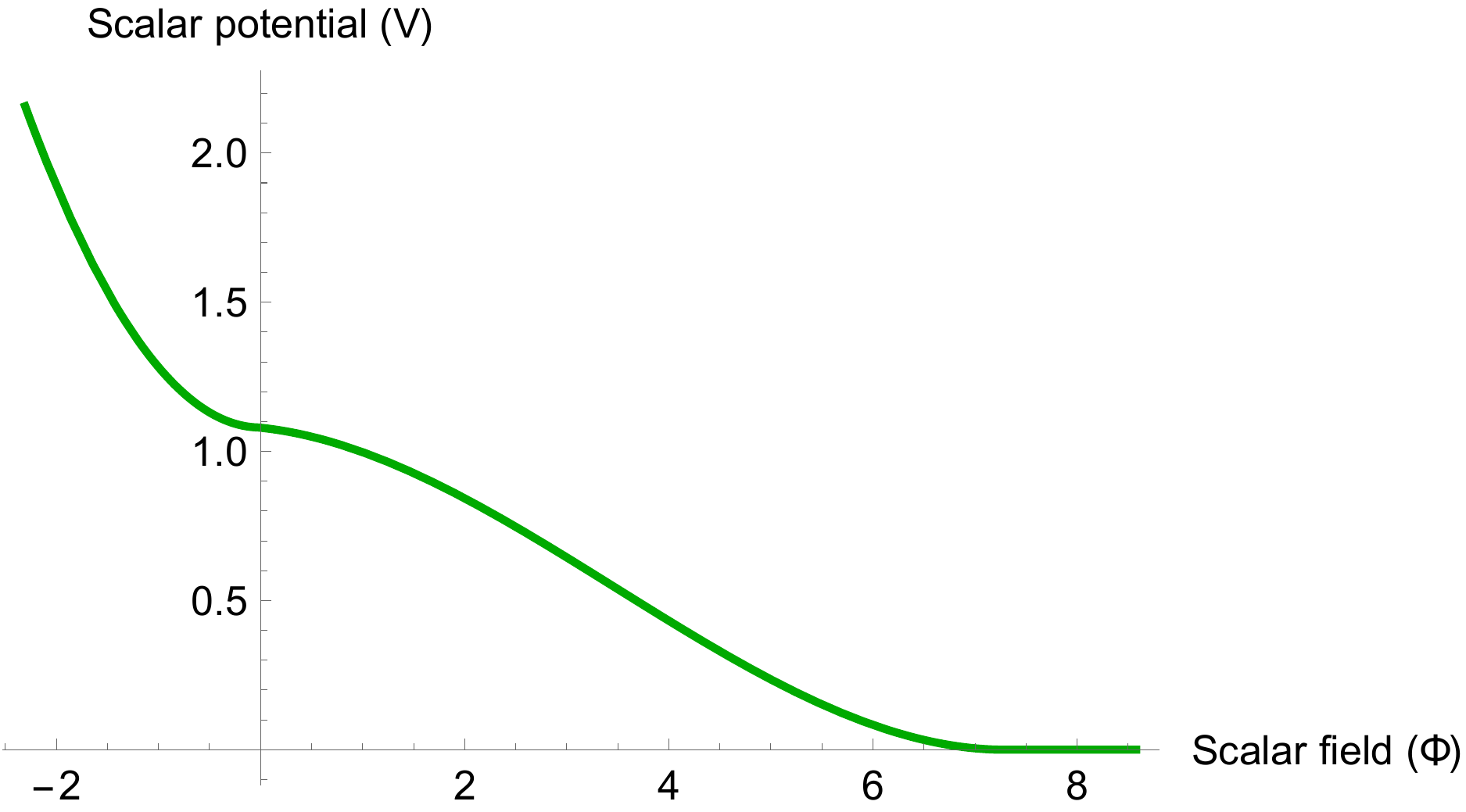}
\caption{Scalar potential $V$ has been depicted against the scalar field $\Phi$. The potential decreases steeply with time and hence the slow-roll approximation will not work in this context. As the inflation ends the potential becomes flat and hence having little influence on dynamics of the universe.}
\label{fig_06}
\end{center}
\end{figure}

Using the form of scalar potential, coupling function and the Hubble parameter one can easily solve for the only remaining bit, i.e., the scalar field equation numerically. Given the scalar field potential as a function of time as well as the scalar field as a function of time one can eliminate time from the two and hence plot the potential as a function of the scalar field. This is what we have presented in \ref{fig_06}, where the scalar field as well as the potential have been ``rescaled'' in an appropriate manner.

In order to match the numerical solution for the scalar field with the analytic ones, we use suitable boundary conditions on $\Phi$ and $\dot{\Phi}$ respectively. From \ref{fig_06}, it is clear that the scalar field rolls down the scalar potential $V(\Phi)$ in a rapid manner and hence it is completely consistent with our earlier findings that slow-roll approximations will not work here. Finally for $t>6\times 10^{-8}~\textrm{GeV}^{-1}$, the potential becomes flat and the field exits from inflation. This is completely consistent with our analytical estimates as well. Thus from \ref{scalar_solution_1} and \ref{sol_phi_h_02}, one can easily conclude that just like the Hubble parameter, the numerical solution of scalar field also matches with the analytic one near about the beginning and the end stages of inflation.

\begin{figure}[htp]
\begin{center}
\centering
\includegraphics[scale=0.6]{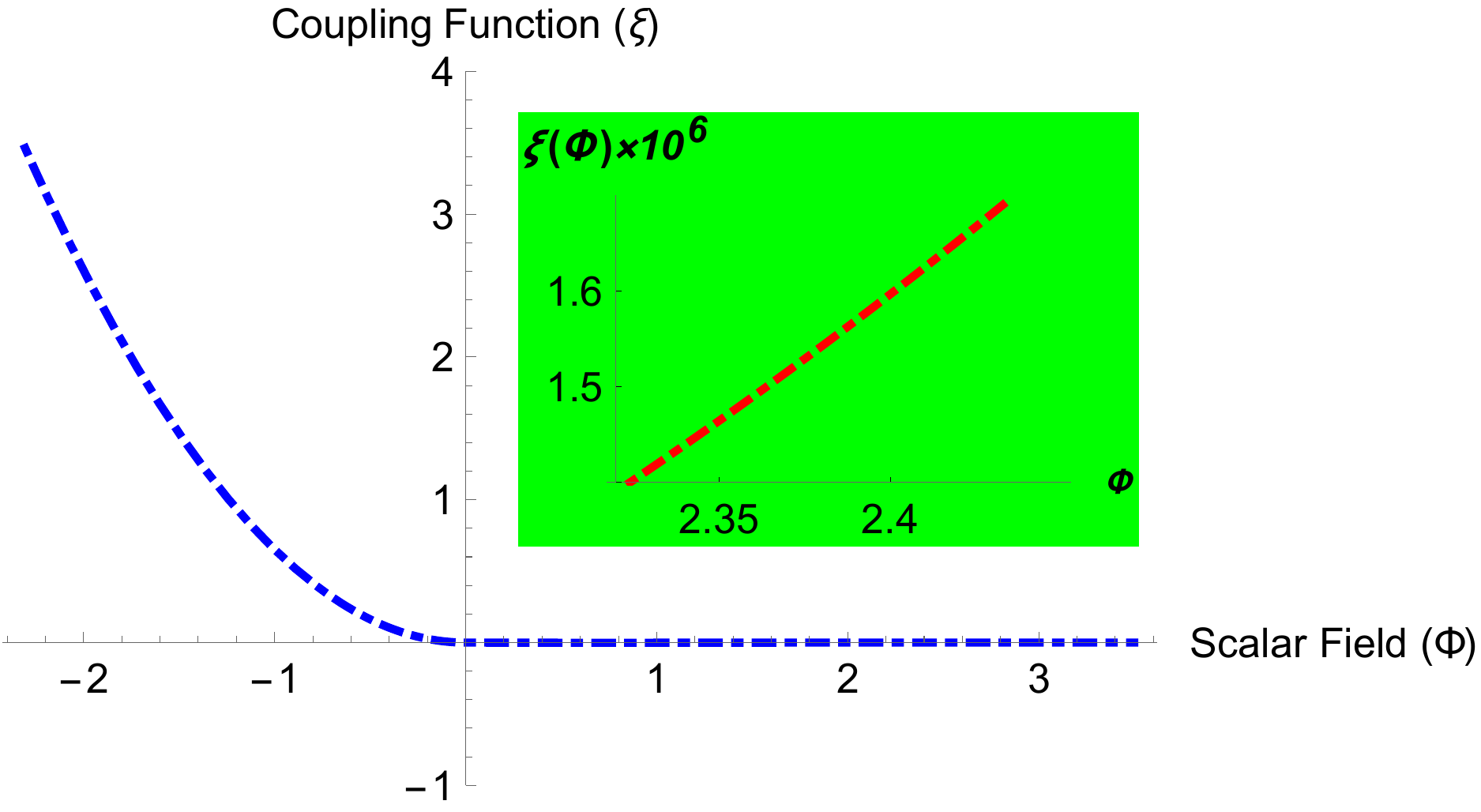}
\caption{The coupling function $\xi$ has been presented against the scalar field $\Phi$. As evident the coupling function decreases steeply as the scalar field reaches larger and larger values. Hence the slow-roll approximation for the scalar field will not work in this context. As the inflation ends the potential indeed increases (see the graph in the inset), but is very small in magnitude and hence have very little influence on the post-inflationary dynamics of the universe.}
\label{fig_07}
\end{center}
\end{figure}

For completeness, we have also presented variation of the coupling function $\xi(\Phi)$ with the scalar field $\Phi$. As expected it presents a rapid fall in the initial stages of inflation and becomes very small near the end of the inflation (see \ref{fig_07}), after which it again starts to increase (see the inset figure of \ref{fig_07}). However the numerical value of the coupling function during this late time increment is very small and hence one can safely argue that after exit from the inflationary scenario the \GB term will have little influence on the dynamics of the universe. As a consequence the ratio $(\xi(t)\mathcal{G}/8\pi G R)\sim \mathcal{O}(10^{-27})$ just after the end of the inflation. Thus once the universe exits from inflationary period, the Gauss-Bonnet term (coupled with the scalar field) can be safely ignored with respect to the Ricci scalar and hence the universe is dominated only by Einstein's gravity.

Finally, let us briefly comment on possible observational signatures of the model under consideration. In the context of inflationary paradigm the key observational parameters are the tensor to scalar ratio $r$ and the scalar spectral index $n_{s}$. Both of which have been computed in \ref{Inf_self_int} and similar numerical values for these two observational parameters also hold for the present situation as well. Both of these values are well within the observational bounds advocated by the Planck mission and hence are consistent with the current observational estimations. There are several other possibilities, where the observational feasibility of this model can be commented upon or some forecast can be provided, which later on can be verified. For example, an estimation of the three point correlation function, which in turn is related to the non-Gaussianity parameter, may lead to some non-trivial results over and above the standard inflationary background. Furthermore, the effect of the non-trivial coupling between the inflaton field and the \GB invariant may lead to interesting implications for polarization modes of the photons originating from the last scattering surface. These issues deserve further investigation, which we leave for the future.

\section{Concluding Remarks}

In this work we set out to explore the influence of the \GB term on the inflationary paradigm. In particular, even though the \GB term alone in four spacetime dimensions is topological in nature, a non-trivial coupling of the same with the inflaton field can influence the evolution of the universe. To understand the effect of the coupling of the \GB term in some detail we consider a particular scenario in which the self-interacting potential for the inflaton field is absent. By solving the associated field equations we could explicitly show that the above model indeed exhibits an exponential expansion of the universe. Subsequently, using the reconstruction technique, we have been able to argue that the \GB term coupled with a scalar field can indeed drive the inflation of the universe, while also providing an exit. The above model turned out to be consistent with current observations as well. However, the scalar coupled \GB term encounters difficulty when one considers evolution of tensor perturbations and in general circumstances we have been able to demonstrate that it will always be unstable. This motivates us to introduce the self-interacting potential for the scalar field. Unlike the results derived in earlier literatures, here we have considered the effect of the \GB invariant as well as the scalar potential on the inflationary paradigm. Having derived the initial accelerating phase and the final decelerating phase we have interpolated the behaviour of the Hubble parameter, the scalar field and the potential between these two phases numerically. It turns out that in both these contexts, with or without the potential, the scalar coupling to the \GB term gradually decreases to small and constant value as the universe exits from inflation. Thus \emph{after} the universe exits from inflation, the \GB term has negligible influence on the dynamics of the universe. Hence as the inflation ends the scalar coupled \GB term goes out of the dynamical picture, such that afterwards the evolution of the universe is governed by the Einstein term alone. 
\section*{Acknowledgements}

Research of SC is funded by the INSPIRE Faculty fellowship (Reg. No. DST/INSPIRE/04/2018/000893)
from Department of Science and Technology, Government of India. The research of SSG is supported by the Science and Engineering Research Board-Extra Mural Research Grant (No. EMR/2017/001372), Government of India. Finally, SC would like to thank Albert Einstein Institute, Potsdam, Germany for warm hospitality, where a part of this work was completed. 
\bibliography{References}

\providecommand{\href}[2]{#2}\begingroup\raggedright\begin{thebibliography}{100}

\bibitem{will1993theory}
C.~M. Will, {\em Theory and experiment in gravitational physics}.
\newblock Cambridge University Press, 1993.

\bibitem{carroll2004spacetime}
S.~M. Carroll, {\em Spacetime and geometry. An introduction to general
  relativity}, vol.~1.
\newblock 2004.

\bibitem{gravitation}
T.Padmanabhan, {\em {Gravitation: Foundations and Frontiers}}.
\newblock Cambridge University Press, Cambridge, UK, 2010.

\bibitem{MTW}
C.~W. Misner, K.~S. Thorne, and J.~A. Wheeler, {\em {Gravitation}}.
\newblock W. H. Freeman and Company, 3~ed., 1973.

\bibitem{Guth:1980zm}
A.~H. Guth, ``{The Inflationary Universe: A Possible Solution to the Horizon
  and Flatness Problems},''
\href{http://dx.doi.org/10.1103/PhysRevD.23.347}{{\em Phys. Rev.} {\bfseries
  D23} (1981) 347--356}.

\bibitem{Starobinsky:1980te}
A.~A. Starobinsky, ``{A New Type of Isotropic Cosmological Models Without
  Singularity},''
\href{http://dx.doi.org/10.1016/0370-2693(80)90670-X}{{\em Phys. Lett.}
  {\bfseries 91B} (1980) 99--102}.

\bibitem{Linde:1981mu}
A.~D. Linde, ``{A New Inflationary Universe Scenario: A Possible Solution of
  the Horizon, Flatness, Homogeneity, Isotropy and Primordial Monopole
  Problems},''
\href{http://dx.doi.org/10.1016/0370-2693(82)91219-9}{{\em Phys. Lett.}
  {\bfseries 108B} (1982) 389--393}.

\bibitem{Linde:1982zj}
A.~D. Linde, ``{Coleman-Weinberg Theory and a New Inflationary Universe
  Scenario},''
\href{http://dx.doi.org/10.1016/0370-2693(82)90086-7}{{\em Phys. Lett.}
  {\bfseries 114B} (1982) 431--435}.

\bibitem{Riess:1998cb}
{\bfseries Supernova Search Team} Collaboration, A.~G. Riess {\em et~al.},
  ``{Observational evidence from supernovae for an accelerating universe and a
  cosmological constant},'' \href{http://dx.doi.org/10.1086/300499}{{\em
  Astron. J.} {\bfseries 116} (1998) 1009--1038},
\href{http://arxiv.org/abs/astro-ph/9805201}{{\ttfamily arXiv:astro-ph/9805201
  [astro-ph]}}.

\bibitem{Perlmutter:1998np}
{\bfseries Supernova Cosmology Project} Collaboration, S.~Perlmutter {\em
  et~al.}, ``{Measurements of Omega and Lambda from 42 high redshift
  supernovae},'' \href{http://dx.doi.org/10.1086/307221}{{\em Astrophys. J.}
  {\bfseries 517} (1999) 565--586},
\href{http://arxiv.org/abs/astro-ph/9812133}{{\ttfamily arXiv:astro-ph/9812133
  [astro-ph]}}.

\bibitem{Rubakov:1983bz}
V.~Rubakov and M.~Shaposhnikov, ``{Extra Space-Time Dimensions: Towards a
  Solution to the Cosmological Constant Problem},''
\href{http://dx.doi.org/10.1016/0370-2693(83)91254-6}{{\em Phys.Lett.}
  {\bfseries B125} (1983) 139}.

\bibitem{Peebles:2002gy}
P.~J.~E. Peebles and B.~Ratra, ``{The Cosmological constant and dark energy},''
  \href{http://dx.doi.org/10.1103/RevModPhys.75.559}{{\em Rev. Mod. Phys.}
  {\bfseries 75} (2003) 559--606},
\href{http://arxiv.org/abs/astro-ph/0207347}{{\ttfamily arXiv:astro-ph/0207347
  [astro-ph]}}.

\bibitem{Carroll:2000fy}
S.~M. Carroll, ``{The Cosmological constant},''
  \href{http://dx.doi.org/10.12942/lrr-2001-1}{{\em Living Rev. Rel.}
  {\bfseries 4} (2001) 1},
\href{http://arxiv.org/abs/astro-ph/0004075}{{\ttfamily arXiv:astro-ph/0004075
  [astro-ph]}}.

\bibitem{Padmanabhan:2002ji}
T.~Padmanabhan, ``{Cosmological constant: The Weight of the vacuum},''
  \href{http://dx.doi.org/10.1016/S0370-1573(03)00120-0}{{\em Phys. Rept.}
  {\bfseries 380} (2003) 235--320},
\href{http://arxiv.org/abs/hep-th/0212290}{{\ttfamily arXiv:hep-th/0212290
  [hep-th]}}.

\bibitem{Abbott:1982hn}
L.~F. Abbott, E.~Farhi, and M.~B. Wise, ``{Particle Production in the New
  Inflationary Cosmology},''
\href{http://dx.doi.org/10.1016/0370-2693(82)90867-X}{{\em Phys. Lett.}
  {\bfseries 117B} (1982) 29}.

\bibitem{Linde:1983gd}
A.~D. Linde, ``{Chaotic Inflation},''
\href{http://dx.doi.org/10.1016/0370-2693(83)90837-7}{{\em Phys. Lett.}
  {\bfseries 129B} (1983) 177--181}.

\bibitem{Dodelson:2003ft}
S.~Dodelson, {\em {Modern Cosmology}}.
\newblock Academic Press, Amsterdam, 2003.
\newblock
\url{http://www.slac.stanford.edu/spires/find/books/www?cl=QB981:D62:2003}.
\newblock

\bibitem{Turok:2002yq}
N.~Turok, ``{A critical review of inflation},''
\href{http://dx.doi.org/10.1088/0264-9381/19/13/305}{{\em Class. Quant. Grav.}
  {\bfseries 19} (2002) 3449--3467}.

\bibitem{Lyth:1993eu}
D.~H. Lyth, ``{Introduction to cosmology},'' in {\em {Proceedings, Summer
  School in High-energy physics and cosmology: Trieste, Italy, June 14-July 30,
  1993}}, pp.~0069--136.
\newblock 1993.
\newblock
\href{http://arxiv.org/abs/astro-ph/9312022}{{\ttfamily arXiv:astro-ph/9312022
  [astro-ph]}}.
\newblock

\bibitem{Liddle:1999mq}
A.~R. Liddle, ``{An Introduction to cosmological inflation},'' in {\em
  {Proceedings, Summer School in High-energy physics and cosmology: Trieste,
  Italy, June 29-July 17, 1998}}, pp.~260--295.
\newblock 1999.
\newblock
\href{http://arxiv.org/abs/astro-ph/9901124}{{\ttfamily arXiv:astro-ph/9901124
  [astro-ph]}}.
\newblock

\bibitem{Brandenberger:1999sw}
R.~H. Brandenberger, ``{Inflationary cosmology: Progress and problems},'' in
  {\em {IPM School on Cosmology 1999: Large Scale Structure Formation Tehran,
  Iran, January 23-February 4, 1999}}.
\newblock 1999.
\newblock
\href{http://arxiv.org/abs/hep-ph/9910410}{{\ttfamily arXiv:hep-ph/9910410
  [hep-ph]}}.
\newblock

\bibitem{Guth:2000ka}
A.~H. Guth, ``{Inflation and eternal inflation},''
  \href{http://dx.doi.org/10.1016/S0370-1573(00)00037-5}{{\em Phys. Rept.}
  {\bfseries 333} (2000) 555--574},
\href{http://arxiv.org/abs/astro-ph/0002156}{{\ttfamily arXiv:astro-ph/0002156
  [astro-ph]}}.

\bibitem{Lidsey:1995np}
J.~E. Lidsey, A.~R. Liddle, E.~W. Kolb, E.~J. Copeland, T.~Barreiro, and
  M.~Abney, ``{Reconstructing the inflation potential : An overview},''
  \href{http://dx.doi.org/10.1103/RevModPhys.69.373}{{\em Rev. Mod. Phys.}
  {\bfseries 69} (1997) 373--410},
\href{http://arxiv.org/abs/astro-ph/9508078}{{\ttfamily arXiv:astro-ph/9508078
  [astro-ph]}}.

\bibitem{Burgess:2001vr}
C.~P. Burgess, P.~Martineau, F.~Quevedo, G.~Rajesh, and R.~J. Zhang, ``{Brane -
  anti-brane inflation in orbifold and orientifold models},''
  \href{http://dx.doi.org/10.1088/1126-6708/2002/03/052}{{\em JHEP} {\bfseries
  03} (2002) 052},
\href{http://arxiv.org/abs/hep-th/0111025}{{\ttfamily arXiv:hep-th/0111025
  [hep-th]}}.

\bibitem{Padmanabhan:1988ji}
T.~Padmanabhan and T.~R. Seshadri, ``{Does inflation solve the horizon
  problem?},''
\href{http://dx.doi.org/10.1088/0264-9381/5/1/025}{{\em Class. Quant. Grav.}
  {\bfseries 5} (1988) 221--224}.

\bibitem{Padmanabhan:1988se}
T.~Padmanabhan, T.~R. Seshadri, and T.~P. Singh, ``{Making Inflation Work:
  Damping of Density Perturbations Due to Planck Energy Cutoff},''
\href{http://dx.doi.org/10.1103/PhysRevD.39.2100}{{\em Phys. Rev.} {\bfseries
  D39} (1989) 2100}.

\bibitem{Zwiebach:1985uq}
B.~Zwiebach, ``{Curvature Squared Terms and String Theories},''
\href{http://dx.doi.org/10.1016/0370-2693(85)91616-8}{{\em Phys. Lett.}
  {\bfseries 156B} (1985) 315--317}.

\bibitem{Gross:1986mw}
D.~J. Gross and J.~H. Sloan, ``{The Quartic Effective Action for the Heterotic
  String},''
\href{http://dx.doi.org/10.1016/0550-3213(87)90465-2}{{\em Nucl. Phys.}
  {\bfseries B291} (1987) 41--89}.

\bibitem{Dadhich:2008df}
N.~Dadhich, ``{Characterization of the Lovelock gravity by Bianchi
  derivative},'' \href{http://dx.doi.org/10.1007/s12043-010-0080-1}{{\em
  Pramana} {\bfseries 74} (2010) 875--882},
\href{http://arxiv.org/abs/0802.3034}{{\ttfamily arXiv:0802.3034 [gr-qc]}}.

\bibitem{Padmanabhan:2013xyr}
T.~Padmanabhan and D.~Kothawala, ``{Lanczos-Lovelock models of gravity},''
  \href{http://dx.doi.org/10.1016/j.physrep.2013.05.007}{{\em Phys.Rept.}
  {\bfseries 531} (2013) 115--171},
\href{http://arxiv.org/abs/1302.2151}{{\ttfamily arXiv:1302.2151 [gr-qc]}}.

\bibitem{Woodard:2015zca}
R.~P. Woodard, ``{Ostrogradsky's theorem on Hamiltonian instability},''
  \href{http://dx.doi.org/10.4249/scholarpedia.32243}{{\em Scholarpedia}
  {\bfseries 10} no.~8, (2015) 32243},
\href{http://arxiv.org/abs/1506.02210}{{\ttfamily arXiv:1506.02210 [hep-th]}}.

\bibitem{Satoh:2008ck}
M.~Satoh and J.~Soda, ``{Higher Curvature Corrections to Primordial
  Fluctuations in Slow-roll Inflation},''
  \href{http://dx.doi.org/10.1088/1475-7516/2008/09/019}{{\em JCAP} {\bfseries
  0809} (2008) 019},
\href{http://arxiv.org/abs/0806.4594}{{\ttfamily arXiv:0806.4594 [astro-ph]}}.

\bibitem{Guo:2010jr}
Z.-K. Guo and D.~J. Schwarz, ``{Slow-roll inflation with a Gauss-Bonnet
  correction},'' \href{http://dx.doi.org/10.1103/PhysRevD.81.123520}{{\em Phys.
  Rev.} {\bfseries D81} (2010) 123520},
\href{http://arxiv.org/abs/1001.1897}{{\ttfamily arXiv:1001.1897 [hep-th]}}.

\bibitem{Jiang:2013gza}
P.-X. Jiang, J.-W. Hu, and Z.-K. Guo, ``{Inflation coupled to a Gauss-Bonnet
  term},'' \href{http://dx.doi.org/10.1103/PhysRevD.88.123508}{{\em Phys. Rev.}
  {\bfseries D88} (2013) 123508},
\href{http://arxiv.org/abs/1310.5579}{{\ttfamily arXiv:1310.5579 [hep-th]}}.

\bibitem{Koh:2014bka}
S.~Koh, B.-H. Lee, W.~Lee, and G.~Tumurtushaa, ``{Observational constraints on
  slow-roll inflation coupled to a Gauss-Bonnet term},''
  \href{http://dx.doi.org/10.1103/PhysRevD.90.063527}{{\em Phys. Rev.}
  {\bfseries D90} no.~6, (2014) 063527},
\href{http://arxiv.org/abs/1404.6096}{{\ttfamily arXiv:1404.6096 [gr-qc]}}.

\bibitem{Kanti:2015pda}
P.~Kanti, R.~Gannouji, and N.~Dadhich, ``{Gauss-Bonnet Inflation},''
  \href{http://dx.doi.org/10.1103/PhysRevD.92.041302}{{\em Phys. Rev.}
  {\bfseries D92} no.~4, (2015) 041302},
\href{http://arxiv.org/abs/1503.01579}{{\ttfamily arXiv:1503.01579 [hep-th]}}.

\bibitem{Kanti:2015dra}
P.~Kanti, R.~Gannouji, and N.~Dadhich, ``{Early-time cosmological solutions in
  Einstein-scalar-Gauss-Bonnet theory},''
  \href{http://dx.doi.org/10.1103/PhysRevD.92.083524}{{\em Phys. Rev.}
  {\bfseries D92} no.~8, (2015) 083524},
\href{http://arxiv.org/abs/1506.04667}{{\ttfamily arXiv:1506.04667 [hep-th]}}.

\bibitem{Lahiri:2016jqv}
S.~Lahiri, ``{Anisotropic inflation in Gauss-Bonnet gravity},''
  \href{http://dx.doi.org/10.1088/1475-7516/2016/09/025}{{\em JCAP} {\bfseries
  1609} no.~09, (2016) 025},
\href{http://arxiv.org/abs/1605.09247}{{\ttfamily arXiv:1605.09247 [hep-th]}}.

\bibitem{vandeBruck:2016xvt}
C.~van~de Bruck, K.~Dimopoulos, and C.~Longden, ``{Reheating in
  Gauss-Bonnet-coupled inflation},''
  \href{http://dx.doi.org/10.1103/PhysRevD.94.023506}{{\em Phys. Rev.}
  {\bfseries D94} no.~2, (2016) 023506},
\href{http://arxiv.org/abs/1605.06350}{{\ttfamily arXiv:1605.06350
  [astro-ph.CO]}}.

\bibitem{Koh:2016abf}
S.~Koh, B.-H. Lee, and G.~Tumurtushaa, ``{Reconstruction of the Scalar Field
  Potential in Inflationary Models with a Gauss-Bonnet term},''
  \href{http://dx.doi.org/10.1103/PhysRevD.95.123509}{{\em Phys. Rev.}
  {\bfseries D95} no.~12, (2017) 123509},
\href{http://arxiv.org/abs/1610.04360}{{\ttfamily arXiv:1610.04360 [gr-qc]}}.

\bibitem{Sberna:2017xqv}
L.~Sberna and P.~Pani, ``{Nonsingular solutions and instabilities in
  Einstein-scalar-Gauss-Bonnet cosmology},''
  \href{http://dx.doi.org/10.1103/PhysRevD.96.124022}{{\em Phys. Rev.}
  {\bfseries D96} no.~12, (2017) 124022},
\href{http://arxiv.org/abs/1708.06371}{{\ttfamily arXiv:1708.06371 [gr-qc]}}.

\bibitem{Fomin:2017vae}
I.~V. Fomin and S.~V. Chervon, ``{Exact inflation in Einstein–Gauss–Bonnet
  gravity},'' \href{http://dx.doi.org/10.1134/S0202289317040090}{{\em Grav.
  Cosmol.} {\bfseries 23} no.~4, (2017) 367--374},
\href{http://arxiv.org/abs/1704.03634}{{\ttfamily arXiv:1704.03634 [gr-qc]}}.

\bibitem{vandeBruck:2017voa}
C.~van~de Bruck, K.~Dimopoulos, C.~Longden, and C.~Owen,
  ``{Gauss-Bonnet-coupled Quintessential Inflation},''
\href{http://arxiv.org/abs/1707.06839}{{\ttfamily arXiv:1707.06839
  [astro-ph.CO]}}.

\bibitem{Nojiri:2010wj}
S.~Nojiri and S.~D. Odintsov, ``{Unified cosmic history in modified gravity:
  from F(R) theory to Lorentz non-invariant models},''
  \href{http://dx.doi.org/10.1016/j.physrep.2011.04.001}{{\em Phys.Rept.}
  {\bfseries 505} (2011) 59--144},
\href{http://arxiv.org/abs/1011.0544}{{\ttfamily arXiv:1011.0544 [gr-qc]}}.

\bibitem{Sotiriou:2008rp}
T.~P. Sotiriou and V.~Faraoni, ``{f(R) Theories Of Gravity},''
  \href{http://dx.doi.org/10.1103/RevModPhys.82.451}{{\em Rev.Mod.Phys.}
  {\bfseries 82} (2010) 451--497},
\href{http://arxiv.org/abs/0805.1726}{{\ttfamily arXiv:0805.1726 [gr-qc]}}.

\bibitem{DeFelice:2010aj}
A.~De~Felice and S.~Tsujikawa, ``{f(R) theories},''
  \href{http://dx.doi.org/10.12942/lrr-2010-3}{{\em Living Rev.Rel.} {\bfseries
  13} (2010) 3},
\href{http://arxiv.org/abs/1002.4928}{{\ttfamily arXiv:1002.4928 [gr-qc]}}.

\bibitem{Nojiri:2007as}
S.~Nojiri and S.~D. Odintsov, ``{Unifying inflation with LambdaCDM epoch in
  modified f(R) gravity consistent with Solar System tests},''
  \href{http://dx.doi.org/10.1016/j.physletb.2007.10.027}{{\em Phys. Lett.}
  {\bfseries B657} (2007) 238--245},
\href{http://arxiv.org/abs/0707.1941}{{\ttfamily arXiv:0707.1941 [hep-th]}}.

\bibitem{Nojiri:2003ft}
S.~Nojiri and S.~D. Odintsov, ``{Modified gravity with negative and positive
  powers of the curvature: Unification of the inflation and of the cosmic
  acceleration},'' \href{http://dx.doi.org/10.1103/PhysRevD.68.123512}{{\em
  Phys. Rev.} {\bfseries D68} (2003) 123512},
\href{http://arxiv.org/abs/hep-th/0307288}{{\ttfamily arXiv:hep-th/0307288
  [hep-th]}}.

\bibitem{Nojiri:2017ncd}
S.~Nojiri, S.~D. Odintsov, and V.~K. Oikonomou, ``{Modified Gravity Theories on
  a Nutshell: Inflation, Bounce and Late-time Evolution},''
\href{http://arxiv.org/abs/1705.11098}{{\ttfamily arXiv:1705.11098 [gr-qc]}}.

\bibitem{Oikonomou:2018npe}
V.~K. Oikonomou, ``{Exponential Inflation with $F(R)$ Gravity},''
  \href{http://dx.doi.org/10.1103/PhysRevD.97.064001}{{\em Phys. Rev.}
  {\bfseries D97} no.~6, (2018) 064001},
\href{http://arxiv.org/abs/1801.03426}{{\ttfamily arXiv:1801.03426 [gr-qc]}}.

\bibitem{Chakraborty:2014xla}
S.~Chakraborty and S.~SenGupta, ``{Spherically symmetric brane spacetime with
  bulk $f(\mathcal {R})$ gravity},''
  \href{http://dx.doi.org/10.1140/epjc/s10052-014-3234-3}{{\em Eur.Phys.J.}
  {\bfseries C75} no.~1, (2015) 11},
\href{http://arxiv.org/abs/1409.4115}{{\ttfamily arXiv:1409.4115 [gr-qc]}}.

\bibitem{Chakraborty:2015bja}
S.~Chakraborty and S.~SenGupta, ``{Effective gravitational field equations on
  m-brane embedded in n-dimensional bulk of Einstein and f(R) gravity},''
  \href{http://dx.doi.org/10.1140/epjc/s10052-015-3768-z}{{\em Eur. Phys. J.}
  {\bfseries C75} no.~11, (2015) 538},
\href{http://arxiv.org/abs/1504.07519}{{\ttfamily arXiv:1504.07519 [gr-qc]}}.

\bibitem{Capozziello:1996xg}
S.~Capozziello, R.~de~Ritis, and A.~A. Marino, ``{Some aspects of the
  cosmological conformal equivalence between 'Jordan frame' and 'Einstein
  frame'},'' \href{http://dx.doi.org/10.1088/0264-9381/14/12/010}{{\em Class.
  Quant. Grav.} {\bfseries 14} (1997) 3243--3258},
\href{http://arxiv.org/abs/gr-qc/9612053}{{\ttfamily arXiv:gr-qc/9612053
  [gr-qc]}}.

\bibitem{Sotiriou:2006hs}
T.~P. Sotiriou, ``{f(R) gravity and scalar-tensor theory},''
  \href{http://dx.doi.org/10.1088/0264-9381/23/17/003}{{\em Class. Quant.
  Grav.} {\bfseries 23} (2006) 5117--5128},
\href{http://arxiv.org/abs/gr-qc/0604028}{{\ttfamily arXiv:gr-qc/0604028
  [gr-qc]}}.

\bibitem{Catena:2006bd}
R.~Catena, M.~Pietroni, and L.~Scarabello, ``{Einstein and Jordan reconciled: a
  frame-invariant approach to scalar-tensor cosmology},''
  \href{http://dx.doi.org/10.1103/PhysRevD.76.084039}{{\em Phys. Rev.}
  {\bfseries D76} (2007) 084039},
\href{http://arxiv.org/abs/astro-ph/0604492}{{\ttfamily arXiv:astro-ph/0604492
  [astro-ph]}}.

\bibitem{Chakraborty:2016ydo}
S.~Chakraborty and S.~SenGupta, ``{Solving higher curvature gravity
  theories},'' \href{http://dx.doi.org/10.1140/epjc/s10052-016-4394-0}{{\em
  Eur. Phys. J.} {\bfseries C76} no.~10, (2016) 552},
\href{http://arxiv.org/abs/1604.05301}{{\ttfamily arXiv:1604.05301 [gr-qc]}}.

\bibitem{Chakraborty:2016gpg}
S.~Chakraborty and S.~SenGupta, ``{Gravity stabilizes itself},''
  \href{http://dx.doi.org/10.1140/epjc/s10052-017-5138-5}{{\em Eur. Phys. J.}
  {\bfseries C77} (2017) 573},
\href{http://arxiv.org/abs/1701.01032}{{\ttfamily arXiv:1701.01032 [gr-qc]}}.

\bibitem{Paul:2018kdq}
T.~Paul and S.~Sengupta, ``{Radion tunneling in higher curvature gravity},''
\href{http://arxiv.org/abs/1801.05027}{{\ttfamily arXiv:1801.05027 [hep-th]}}.

\bibitem{Karam:2018squ}
A.~Karam, A.~Lykkas, and K.~Tamvakis, ``{Frame-invariant approach to
  higher-dimensional scalar-tensor gravity},''
\href{http://arxiv.org/abs/1803.04960}{{\ttfamily arXiv:1803.04960 [gr-qc]}}.

\bibitem{Sami:2017nhw}
H.~Sami, J.~Ntahompagaze, and A.~Abebe, ``{Inflationary $f(R)$ Cosmologies},''
  \href{http://dx.doi.org/10.3390/universe3040073}{{\em Universe} {\bfseries 3}
  no.~4, (2017) 73},
\href{http://arxiv.org/abs/1709.04860}{{\ttfamily arXiv:1709.04860 [gr-qc]}}.

\bibitem{Barrow:1988}
J.~D. Barrow and S.~Cotsakis, ``Inflation and the conformal structure of
  higher-order gravity theories,''
  \href{http://dx.doi.org/https://doi.org/10.1016/0370-2693(88)90110-4}{{\em
  Physics Letters B} {\bfseries 214} no.~4, (1988) 515 -- 518}.

\bibitem{Ellis:1991}
G.~F.~R. Ellis and M.~S. Madsen, ``{Exact scalar field cosmologies},'' {\em
  Classical and Quantum Gravity} {\bfseries 8} no.~4, (1991) 667.
  \url{http://stacks.iop.org/0264-9381/8/i=4/a=012}.

\bibitem{Cognola:2005de}
G.~Cognola, E.~Elizalde, S.~Nojiri, S.~D. Odintsov, and S.~Zerbini, ``{One-loop
  f(R) gravity in de Sitter universe},''
  \href{http://dx.doi.org/10.1088/1475-7516/2005/02/010}{{\em JCAP} {\bfseries
  0502} (2005) 010},
\href{http://arxiv.org/abs/hep-th/0501096}{{\ttfamily arXiv:hep-th/0501096
  [hep-th]}}.

\bibitem{Bamba:2008ja}
K.~Bamba and S.~D. Odintsov, ``{Inflation and late-time cosmic acceleration in
  non-minimal Maxwell-$F(R)$ gravity and the generation of large-scale magnetic
  fields},'' \href{http://dx.doi.org/10.1088/1475-7516/2008/04/024}{{\em JCAP}
  {\bfseries 0804} (2008) 024},
\href{http://arxiv.org/abs/0801.0954}{{\ttfamily arXiv:0801.0954 [astro-ph]}}.

\bibitem{Appleby:2009uf}
S.~A. Appleby, R.~A. Battye, and A.~A. Starobinsky, ``{Curing singularities in
  cosmological evolution of F(R) gravity},''
  \href{http://dx.doi.org/10.1088/1475-7516/2010/06/005}{{\em JCAP} {\bfseries
  1006} (2010) 005},
\href{http://arxiv.org/abs/0909.1737}{{\ttfamily arXiv:0909.1737
  [astro-ph.CO]}}.

\bibitem{Sebastiani:2015kfa}
L.~Sebastiani and R.~Myrzakulov, ``{F(R) gravity and inflation},''
  \href{http://dx.doi.org/10.1142/S0219887815300032}{{\em Int. J. Geom. Meth.
  Mod. Phys.} {\bfseries 12} no.~9, (2015) 1530003},
\href{http://arxiv.org/abs/1506.05330}{{\ttfamily arXiv:1506.05330 [gr-qc]}}.

\bibitem{Banerjee:2017lxi}
N.~Banerjee and T.~Paul, ``{Inflationary scenario from higher curvature warped
  spacetime},'' \href{http://dx.doi.org/10.1140/epjc/s10052-017-5256-0}{{\em
  Eur. Phys. J.} {\bfseries C77} no.~10, (2017) 672},
\href{http://arxiv.org/abs/1706.05964}{{\ttfamily arXiv:1706.05964 [hep-th]}}.

\bibitem{Das:2017jrl}
A.~Das, D.~Maity, T.~Paul, and S.~SenGupta, ``{Bouncing cosmology from warped
  extra dimensional scenario},''
  \href{http://dx.doi.org/10.1140/epjc/s10052-017-5396-2}{{\em Eur. Phys. J.}
  {\bfseries C77} no.~12, (2017) 813},
\href{http://arxiv.org/abs/1706.00950}{{\ttfamily arXiv:1706.00950 [hep-th]}}.

\bibitem{Chakraborty:2015wma}
S.~Chakraborty, ``{Lanczos-Lovelock gravity from a thermodynamic
  perspective},'' \href{http://dx.doi.org/10.1007/JHEP08(2015)029}{{\em JHEP}
  {\bfseries 08} (2015) 029},
\href{http://arxiv.org/abs/1505.07272}{{\ttfamily arXiv:1505.07272 [gr-qc]}}.

\bibitem{Chakraborty:2015taq}
S.~Chakraborty and S.~SenGupta, ``{Spherically symmetric brane in a bulk of
  f(R) and Gauss-Bonnet Gravity},''
  \href{http://dx.doi.org/10.1088/0264-9381/33/22/225001}{{\em Class. Quant.
  Grav.} {\bfseries 33} no.~22, (2016) 225001},
\href{http://arxiv.org/abs/1510.01953}{{\ttfamily arXiv:1510.01953 [gr-qc]}}.

\bibitem{Chakraborty:2017zep}
S.~Chakraborty, K.~Parattu, and T.~Padmanabhan, ``{A Novel Derivation of the
  Boundary Term for the Action in Lanczos-Lovelock Gravity},''
  \href{http://dx.doi.org/10.1007/s10714-017-2289-5}{{\em Gen. Rel. Grav.}
  {\bfseries 49} no.~9, (2017) 121},
\href{http://arxiv.org/abs/1703.00624}{{\ttfamily arXiv:1703.00624 [gr-qc]}}.

\bibitem{Cognola:2006eg}
G.~Cognola, E.~Elizalde, S.~Nojiri, S.~D. Odintsov, and S.~Zerbini, ``{Dark
  energy in modified Gauss-Bonnet gravity: Late-time acceleration and the
  hierarchy problem},''
  \href{http://dx.doi.org/10.1103/PhysRevD.73.084007}{{\em Phys. Rev.}
  {\bfseries D73} (2006) 084007},
\href{http://arxiv.org/abs/hep-th/0601008}{{\ttfamily arXiv:hep-th/0601008
  [hep-th]}}.

\bibitem{Nojiri:2005jg}
S.~Nojiri and S.~D. Odintsov, ``{Modified Gauss-Bonnet theory as gravitational
  alternative for dark energy},''
  \href{http://dx.doi.org/10.1016/j.physletb.2005.10.010}{{\em Phys. Lett.}
  {\bfseries B631} (2005) 1--6},
\href{http://arxiv.org/abs/hep-th/0508049}{{\ttfamily arXiv:hep-th/0508049
  [hep-th]}}.

\bibitem{Antoniadis:1993jc}
I.~Antoniadis, J.~Rizos, and K.~Tamvakis, ``{Singularity - free cosmological
  solutions of the superstring effective action},''
  \href{http://dx.doi.org/10.1016/0550-3213(94)90120-1}{{\em Nucl. Phys.}
  {\bfseries B415} (1994) 497--514},
\href{http://arxiv.org/abs/hep-th/9305025}{{\ttfamily arXiv:hep-th/9305025
  [hep-th]}}.

\bibitem{Kanti:1995vq}
P.~Kanti, N.~E. Mavromatos, J.~Rizos, K.~Tamvakis, and E.~Winstanley,
  ``{Dilatonic black holes in higher curvature string gravity},''
  \href{http://dx.doi.org/10.1103/PhysRevD.54.5049}{{\em Phys. Rev.} {\bfseries
  D54} (1996) 5049--5058},
\href{http://arxiv.org/abs/hep-th/9511071}{{\ttfamily arXiv:hep-th/9511071
  [hep-th]}}.

\bibitem{Kanti:1997br}
P.~Kanti, N.~E. Mavromatos, J.~Rizos, K.~Tamvakis, and E.~Winstanley,
  ``{Dilatonic black holes in higher curvature string gravity. 2: Linear
  stability},'' \href{http://dx.doi.org/10.1103/PhysRevD.57.6255}{{\em Phys.
  Rev.} {\bfseries D57} (1998) 6255--6264},
\href{http://arxiv.org/abs/hep-th/9703192}{{\ttfamily arXiv:hep-th/9703192
  [hep-th]}}.

\bibitem{Charmousis:2002rc}
C.~Charmousis and J.-F. Dufaux, ``{General Gauss-Bonnet brane cosmology},''
  \href{http://dx.doi.org/10.1088/0264-9381/19/18/304}{{\em Class. Quant.
  Grav.} {\bfseries 19} (2002) 4671--4682},
\href{http://arxiv.org/abs/hep-th/0202107}{{\ttfamily arXiv:hep-th/0202107
  [hep-th]}}.

\bibitem{Binetruy:2002ck}
P.~Binetruy, C.~Charmousis, S.~C. Davis, and J.-F. Dufaux, ``{Avoidance of
  naked singularities in dilatonic brane world scenarios with a Gauss-Bonnet
  term},'' \href{http://dx.doi.org/10.1016/S0370-2693(02)02477-2}{{\em Phys.
  Lett.} {\bfseries B544} (2002) 183--191},
\href{http://arxiv.org/abs/hep-th/0206089}{{\ttfamily arXiv:hep-th/0206089
  [hep-th]}}.

\bibitem{Germani:2002pt}
C.~Germani and C.~F. Sopuerta, ``{String inspired brane world cosmology},''
  \href{http://dx.doi.org/10.1103/PhysRevLett.88.231101}{{\em Phys. Rev. Lett.}
  {\bfseries 88} (2002) 231101},
\href{http://arxiv.org/abs/hep-th/0202060}{{\ttfamily arXiv:hep-th/0202060
  [hep-th]}}.

\bibitem{Gravanis:2002wy}
E.~Gravanis and S.~Willison, ``{Israel conditions for the Gauss-Bonnet theory
  and the Friedmann equation on the brane universe},''
  \href{http://dx.doi.org/10.1016/S0370-2693(03)00555-0}{{\em Phys. Lett.}
  {\bfseries B562} (2003) 118--126},
\href{http://arxiv.org/abs/hep-th/0209076}{{\ttfamily arXiv:hep-th/0209076
  [hep-th]}}.

\bibitem{Nojiri:2005am}
S.~Nojiri, S.~D. Odintsov, and O.~G. Gorbunova, ``{Dark energy problem: From
  phantom theory to modified Gauss-Bonnet gravity},''
  \href{http://dx.doi.org/10.1088/0305-4470/39/21/S62}{{\em J. Phys.}
  {\bfseries A39} (2006) 6627--6634},
\href{http://arxiv.org/abs/hep-th/0510183}{{\ttfamily arXiv:hep-th/0510183
  [hep-th]}}.

\bibitem{Leith:2007bu}
B.~M. Leith and I.~P. Neupane, ``{Gauss-Bonnet cosmologies: Crossing the
  phantom divide and the transition from matter dominance to dark energy},''
  \href{http://dx.doi.org/10.1088/1475-7516/2007/05/019}{{\em JCAP} {\bfseries
  0705} (2007) 019},
\href{http://arxiv.org/abs/hep-th/0702002}{{\ttfamily arXiv:hep-th/0702002
  [hep-th]}}.

\bibitem{Deser:2007jk}
S.~Deser and R.~P. Woodard, ``{Nonlocal Cosmology},''
  \href{http://dx.doi.org/10.1103/PhysRevLett.99.111301}{{\em Phys. Rev. Lett.}
  {\bfseries 99} (2007) 111301},
\href{http://arxiv.org/abs/0706.2151}{{\ttfamily arXiv:0706.2151 [astro-ph]}}.

\bibitem{Bamba:2014mya}
K.~Bamba, A.~N. Makarenko, A.~N. Myagky, and S.~D. Odintsov, ``{Bouncing
  cosmology in modified Gauss-Bonnet gravity},''
  \href{http://dx.doi.org/10.1016/j.physletb.2014.04.004}{{\em Phys. Lett.}
  {\bfseries B732} (2014) 349--355},
\href{http://arxiv.org/abs/1403.3242}{{\ttfamily arXiv:1403.3242 [hep-th]}}.

\bibitem{vandeBruck:2015xpa}
C.~van~de Bruck and L.~E. Paduraru, ``{Simplest extension of Starobinsky
  inflation},'' \href{http://dx.doi.org/10.1103/PhysRevD.92.083513}{{\em Phys.
  Rev.} {\bfseries D92} (2015) 083513},
\href{http://arxiv.org/abs/1505.01727}{{\ttfamily arXiv:1505.01727 [hep-th]}}.

\bibitem{Sotiriou:2013qea}
T.~P. Sotiriou and S.-Y. Zhou, ``{Black hole hair in generalized scalar-tensor
  gravity},'' \href{http://dx.doi.org/10.1103/PhysRevLett.112.251102}{{\em
  Phys. Rev. Lett.} {\bfseries 112} (2014) 251102},
\href{http://arxiv.org/abs/1312.3622}{{\ttfamily arXiv:1312.3622 [gr-qc]}}.

\bibitem{Hees:2017aal}
A.~Hees {\em et~al.}, ``{Testing General Relativity with stellar orbits around
  the supermassive black hole in our Galactic center},''
  \href{http://dx.doi.org/10.1103/PhysRevLett.118.211101}{{\em Phys. Rev.
  Lett.} {\bfseries 118} no.~21, (2017) 211101},
\href{http://arxiv.org/abs/1705.07902}{{\ttfamily arXiv:1705.07902
  [astro-ph.GA]}}.

\bibitem{Antoniou:2017acq}
G.~Antoniou, A.~Bakopoulos, and P.~Kanti, ``{Evasion of No-Hair Theorems and
  Novel Black-Hole Solutions in Gauss-Bonnet Theories},''
  \href{http://dx.doi.org/10.1103/PhysRevLett.120.131102}{{\em Phys. Rev.
  Lett.} {\bfseries 120} no.~13, (2018) 131102},
\href{http://arxiv.org/abs/1711.03390}{{\ttfamily arXiv:1711.03390 [hep-th]}}.

\bibitem{Charmousis:2014mia}
C.~Charmousis, ``{From Lovelock to Horndeski`s Generalized Scalar Tensor
  Theory},'' \href{http://dx.doi.org/10.1007/978-3-319-10070-8_2}{{\em Lect.
  Notes Phys.} {\bfseries 892} (2015) 25--56},
\href{http://arxiv.org/abs/1405.1612}{{\ttfamily arXiv:1405.1612 [gr-qc]}}.

\bibitem{Chakraborty:2016lxo}
S.~Chakraborty and S.~SenGupta, ``{Strong gravitational lensing --- A probe for
  extra dimensions and Kalb-Ramond field},''
  \href{http://dx.doi.org/10.1088/1475-7516/2017/07/045}{{\em JCAP} {\bfseries
  1707} no.~07, (2017) 045},
\href{http://arxiv.org/abs/1611.06936}{{\ttfamily arXiv:1611.06936 [gr-qc]}}.

\bibitem{Banerjee:2017hzw}
I.~Banerjee, S.~Chakraborty, and S.~SenGupta, ``{Excavating black hole
  continuum spectrum: Possible signatures of scalar hairs and of higher
  dimensions},'' \href{http://dx.doi.org/10.1103/PhysRevD.96.084035}{{\em Phys.
  Rev.} {\bfseries D96} no.~8, (2017) 084035},
\href{http://arxiv.org/abs/1707.04494}{{\ttfamily arXiv:1707.04494 [gr-qc]}}.

\bibitem{Mukherjee:2017fqz}
S.~Mukherjee and S.~Chakraborty, ``{Horndeski theories confront Gravity Probe
  B},''
\href{http://arxiv.org/abs/1712.00562}{{\ttfamily arXiv:1712.00562 [gr-qc]}}.

\bibitem{Pirtskhalava:2015nla}
D.~Pirtskhalava, L.~Santoni, E.~Trincherini, and F.~Vernizzi, ``{Weakly Broken
  Galileon Symmetry},''
  \href{http://dx.doi.org/10.1088/1475-7516/2015/09/007}{{\em JCAP} {\bfseries
  1509} no.~09, (2015) 007},
\href{http://arxiv.org/abs/1505.00007}{{\ttfamily arXiv:1505.00007 [hep-th]}}.

\bibitem{Banerjee:2016hom}
S.~Banerjee and E.~N. Saridakis, ``{Bounce and cyclic cosmology in weakly
  broken galileon theories},''
  \href{http://dx.doi.org/10.1103/PhysRevD.95.063523}{{\em Phys. Rev.}
  {\bfseries D95} no.~6, (2017) 063523},
\href{http://arxiv.org/abs/1604.06932}{{\ttfamily arXiv:1604.06932 [gr-qc]}}.

\bibitem{Banerjee:2017jyb}
R.~Banerjee, S.~Chakraborty, A.~Mitra, and P.~Mukherjee, ``{Cosmological
  implications of shift symmetric Galileon field},''
  \href{http://dx.doi.org/10.1103/PhysRevD.96.064023}{{\em Phys. Rev.}
  {\bfseries D96} no.~6, (2017) 064023},
\href{http://arxiv.org/abs/1705.06941}{{\ttfamily arXiv:1705.06941 [gr-qc]}}.

\bibitem{Bhattacharya:2016naa}
S.~Bhattacharya and S.~Chakraborty, ``{Constraining some Horndeski gravity
  theories},'' \href{http://dx.doi.org/10.1103/PhysRevD.95.044037}{{\em Phys.
  Rev.} {\bfseries D95} no.~4, (2017) 044037},
\href{http://arxiv.org/abs/1607.03693}{{\ttfamily arXiv:1607.03693 [gr-qc]}}.

\bibitem{Banerjee:2018yyi}
R.~Banerjee, S.~Chakraborty, and P.~Mukherjee, ``{Late-time acceleration driven
  by shift-symmetric Galileon in presence of Torsion},''
\href{http://arxiv.org/abs/1802.04150}{{\ttfamily arXiv:1802.04150 [gr-qc]}}.

\bibitem{Aghanim:2015xee}
{\bfseries Planck} Collaboration, N.~Aghanim {\em et~al.}, ``{Planck 2015
  results. XI. CMB power spectra, likelihoods, and robustness of parameters},''
  \href{http://dx.doi.org/10.1051/0004-6361/201526926}{{\em Astron. Astrophys.}
  {\bfseries 594} (2016) A11},
\href{http://arxiv.org/abs/1507.02704}{{\ttfamily arXiv:1507.02704
  [astro-ph.CO]}}.

\bibitem{Ade:2015xua}
{\bfseries Planck} Collaboration, P.~A.~R. Ade {\em et~al.}, ``{Planck 2015
  results. XIII. Cosmological parameters},''
  \href{http://dx.doi.org/10.1051/0004-6361/201525830}{{\em Astron. Astrophys.}
  {\bfseries 594} (2016) A13},
\href{http://arxiv.org/abs/1502.01589}{{\ttfamily arXiv:1502.01589
  [astro-ph.CO]}}.

\bibitem{Amendola:2005cr}
L.~Amendola, C.~Charmousis, and S.~C. Davis, ``{Constraints on Gauss-Bonnet
  gravity in dark energy cosmologies},''
  \href{http://dx.doi.org/10.1088/1475-7516/2006/12/020}{{\em JCAP} {\bfseries
  0612} (2006) 020},
\href{http://arxiv.org/abs/hep-th/0506137}{{\ttfamily arXiv:hep-th/0506137
  [hep-th]}}.

\bibitem{Carloni:2010ph}
S.~Carloni, R.~Goswami, and P.~K.~S. Dunsby, ``{A new approach to
  reconstruction methods in $f(R)$ gravity},''
  \href{http://dx.doi.org/10.1088/0264-9381/29/13/135012}{{\em Class. Quant.
  Grav.} {\bfseries 29} (2012) 135012},
\href{http://arxiv.org/abs/1005.1840}{{\ttfamily arXiv:1005.1840 [gr-qc]}}.

\bibitem{Nojiri:2009xh}
S.~Nojiri, S.~D. Odintsov, A.~Toporensky, and P.~Tretyakov, ``{Reconstruction
  and deceleration-acceleration transitions in modified gravity},''
  \href{http://dx.doi.org/10.1007/s10714-010-0977-5}{{\em Gen. Rel. Grav.}
  {\bfseries 42} (2010) 1997--2008},
\href{http://arxiv.org/abs/0912.2488}{{\ttfamily arXiv:0912.2488 [hep-th]}}.

\bibitem{Nojiri:2009kx}
S.~Nojiri, S.~D. Odintsov, and D.~Saez-Gomez, ``{Cosmological reconstruction of
  realistic modified F(R) gravities},''
  \href{http://dx.doi.org/10.1016/j.physletb.2009.09.045}{{\em Phys. Lett.}
  {\bfseries B681} (2009) 74--80},
\href{http://arxiv.org/abs/0908.1269}{{\ttfamily arXiv:0908.1269 [hep-th]}}.

\bibitem{Sberna:2017nzp}
L.~Sberna, ``{Early-universe cosmology in Einstein-scalar-Gauss-Bonnet
  gravity},'' Master's thesis, Rome U., 2017.
\newblock
  \url{http://inspirehep.net/record/1614305/files/arXiv:1708.01150.pdf}.

\bibitem{Kawai:1998ab}
S.~Kawai, M.-a. Sakagami, and J.~Soda, ``{Instability of one loop superstring
  cosmology},'' \href{http://dx.doi.org/10.1016/S0370-2693(98)00925-3}{{\em
  Phys. Lett.} {\bfseries B437} (1998) 284--290},
\href{http://arxiv.org/abs/gr-qc/9802033}{{\ttfamily arXiv:gr-qc/9802033
  [gr-qc]}}.

\bibitem{Kawai:1999pw}
S.~Kawai and J.~Soda, ``{Evolution of fluctuations during graceful exit in
  string cosmology},''
  \href{http://dx.doi.org/10.1016/S0370-2693(99)00736-4}{{\em Phys. Lett.}
  {\bfseries B460} (1999) 41--46},
\href{http://arxiv.org/abs/gr-qc/9903017}{{\ttfamily arXiv:gr-qc/9903017
  [gr-qc]}}.

\bibitem{Hikmawan:2015rze}
G.~Hikmawan, J.~Soda, A.~Suroso, and F.~P. Zen, ``{Comment on “Gauss-Bonnet
  inflation”},'' \href{http://dx.doi.org/10.1103/PhysRevD.93.068301}{{\em
  Phys. Rev.} {\bfseries D93} no.~6, (2016) 068301},
\href{http://arxiv.org/abs/1512.00222}{{\ttfamily arXiv:1512.00222 [hep-th]}}.

\bibitem{Kawai:1997mf}
S.~Kawai, M.-a. Sakagami, and J.~Soda, ``{Perturbative analysis of nonsingular
  cosmological model},'' in {\em {Proceedings, 7th Workshop on General
  Relativity and Gravitation (JGRG7): Kyoto, Japan, October 27-30, 1997}}.
\newblock 1997.
\newblock
\href{http://arxiv.org/abs/gr-qc/9901065}{{\ttfamily arXiv:gr-qc/9901065
  [gr-qc]}}.
\newblock

\bibitem{Cognola:2006sp}
G.~Cognola, E.~Elizalde, S.~Nojiri, S.~Odintsov, and S.~Zerbini,
  ``{String-inspired Gauss-Bonnet gravity reconstructed from the universe
  expansion history and yielding the transition from matter dominance to dark
  energy},'' \href{http://dx.doi.org/10.1103/PhysRevD.75.086002}{{\em Phys.
  Rev.} {\bfseries D75} (2007) 086002},
\href{http://arxiv.org/abs/hep-th/0611198}{{\ttfamily arXiv:hep-th/0611198
  [hep-th]}}.

\end{thebibliography}\endgroup

\bibliographystyle{./utphys1}

\end{document}